\documentclass{article} 
 
\usepackage{fullpage,epsf,overcite} 

\def \bu{ {{\mbox {\boldmath $1$}}}   }
\def \bs{ {{\mbox {\boldmath $\sigma \! $}}}}
\def \btau{ {{\mbox {\boldmath $\tau$}}}}
\def \bgam{ {{\mbox {\boldmath $\gamma$}}}}
\def \beps{ {{\mbox {\boldmath $\epsilon$}}}}
\def \br{ {{\mbox {\boldmath $r$}}}}

\def \bF{ {{\mbox {\boldmath $F$}}}}

\def \bfk{ {{\mbox {\boldmath $k$}}}}
\def \bQ{ {{\mbox {\boldmath $Q$}}}}
\def \bv{ {{\mbox {\boldmath $v$}}}}
\def \bY{ {{\mbox {\boldmath $Y$}}}}
\def \bZ{ {{\mbox {\boldmath $Z$}}}}
\def \bW{ {{\mbox {\boldmath $W$}}}}
\def \bE{ {{\mbox {\boldmath $E$}}}}
\def \bsh{ {\hat \bs}}
\def \bk{ {\mbox {\boldmath $\kappa$}}}
\def \avel { {\bigl\langle}}
\def \aver { {\bigr\rangle}}

\def \bDel{ {\bf \Delta}}
\def \bPi{ {\bf \Pi}}

\def \one{ { { 1 \kern-0.28em {\rm I}}  }}
\def \bsjk{ {\bs_{jk}}}
\def \bshmn{ {\hat \bs_{\mu \nu}}}

\title{\bf {Rouse Chains with Excluded Volume Interactions: 
Linear Viscoelasticity }} 
\author {{\bf J. Ravi Prakash\footnote{Present address: {\it 
Department~of~Chemical~Engineering, Monash University,
Victoria 3800, Australia }}}
\\{\it Department~of~Chemical~Engineering,}\\ {\it 
Indian~Institute~of~Technology, Madras, India 600 036}\\  } 
 
\begin{document} 
\maketitle 
 
\begin{abstract} 

		Linear viscoelastic properties for a dilute polymer
		solution are predicted by modeling the solution as a
		suspension of non-interacting bead-spring chains. The
		present model, unlike the Rouse model, can describe
		the solution's rheological behavior even when the
		solvent quality is {\em good}, since {\em excluded
		volume} effects are explicitly taken into account
		through a narrow Gaussian repulsive potential between
		pairs of beads in a bead-spring chain. The use of the
		narrow Gaussian potential, which tends to the more
		commonly used $\delta$-function repulsive potential in
		the limit of a {\em width} parameter $d$ going to
		zero, enables the performance of Brownian dynamics
		simulations. The simulations results, which describe
		the exact behavior of the model, indicate that for
		chains of {\em arbitrary} but {\em finite} length, a
		$\delta$-function potential leads to equilibrium and
		zero shear rate properties which are identical to the
		predictions of the Rouse model.  On the other hand, a
		non-zero value of $d$ gives rise to a prediction of
		swelling at equilibrium, and an increase in zero shear
		rate properties relative to their Rouse model
		values. The use of a $\delta$-function potential
		appears to be justified in the limit of infinite chain
		length. The exact simulation results are compared with
		those obtained with an approximate solution which is
		based on the assumption that the non-equilibrium
		configurational distribution function is Gaussian. The
		Gaussian approximation is shown to be exact to first
		order in the strength of excluded volume interaction,
		and is found to be accurate above a threshold value of
		$d$, for given values of chain length and strength
		of excluded volume interaction. 

\end{abstract} 
\thispagestyle{empty}
\newpage
\section {\bf {Introduction }} 

The simplest model, within the context of Polymer Kinetic Theory, to
describe the rheological behavior of dilute polymer solutions is the
Rouse model~\cite{rouse}. The Rouse model represents the macromolecule
by a linear chain of identical beads connected by Hookean springs, and
assumes that the solvent influences the motion of the beads by
exerting a drag force and a Brownian force. While the Rouse model is
able to explain the existence of viscoelasticity in polymer solutions
by predicting a constant non-zero first normal stress difference in
simple shear flow, it cannot predict several other features of dilute
solution behavior, such as the existence of a nonzero second normal
stress difference, the existence of shear rate dependent viscometric
functions, or the correct molecular weight dependence of material
functions.  Over the past decade, considerable progress has been made
by incorporating the effect of fluctuating hydrodynamic interaction
into the Rouse model~\cite{ottga,wedgega,zylkaga,prakotthi,prakbk}.
These models are able to predict the molecular weight dependence of
the material functions accurately. They also predict a nonzero second
normal stress difference, and shear rate dependent viscometric
functions.  However, since they neglect the existence of excluded
volume interactions among parts of the polymer chain, they are
strictly applicable only to {\em theta} solutions.

Recently, Prakash and \"Ottinger~\cite{prakottev} examined the
influence of excluded volume effects on the rheological behavior of
dilute polymer solutions by representing the polymer molecule with a
Hookean dumbbell model, and using a narrow Gaussian repulsive
potential to describe the excluded volume interactions between the
beads of the dumbbell.  The narrow Gaussian potential tends to a
$\delta$-function potential in the limit of a parameter, that
describes the width of the potential, going to zero. It can therefore
be used to evaluate results obtained with the singular
$\delta$-function potential.  It was shown by them that the use of a
$\delta$-function potential between the beads, which is commonly used
in static theories for polymer solutions~\cite{doi,declos,schaf},
leads to no change in the equilibrium or dynamic properties of the
dumbbell when compared to the case where no excluded volume
interactions are taken into account. They also found that assuming
that the non-equilibrium configurational distribution function is a
Gaussian leads to accurate predictions of viscometric functions in a
certain range of parameter values. These results suggest that it would
be worthwhile examining longer bead-spring chains. Firstly, it is
interesting to see if the problem with the $\delta$-function potential
can be resolved when there are more beads in the bead-spring
chain. Secondly, it is important to find out if the Gaussian
approximation is accurate even for longer chains.  The purpose of this
paper is to attempt to answer these questions, in the linear
viscoelastic limit, by extending the methodology developed in the
earlier paper to the case of bead-spring chains. The same issues, in
the context of steady shear flows at finite shear rates, will be
addressed in a subsequent paper.

As in the dumbbell paper, we confine our attention to excluded volume
interactions alone, and neglect the presence of hydrodynamic
interactions.  This clearly implies---since it is essential to include
hydrodynamic interaction effects for a proper description of the
dynamic behavior of dilute solutions---that the results of the present
paper are not yet directly comparable with experiments. They represent
a preliminary step in that direction. It is felt that the inclusion of
hydrodynamic interaction would make the theory significantly more
complex before the role of excluded volume interactions is properly
understood. The aim of this work is to develop and carefully evaluate
the Gaussian approximation for excluded volume interactions. The
Gaussian approximation has already been shown to be excellent for the
treatment of hydrodynamic interaction effects~\cite{zylkaga}. If it
also proves to be accurate for the treatment of excluded volume
effects, then it would be extremely useful for describing the combined
effects of hydrodynamic interaction and excluded volume. It should be
noted that, in principle, the development of such an approximation does
not pose any fundamental problems.

This paper is organized as follows. In the next section, the basic
equations governing the dynamics of Rouse chains with excluded volume
interactions are discussed. A retarded motion expansion for the stress
tensor is derived in section~3, and exact expressions for the zero
shear rate viscometric functions in simple shear flow are
obtained. The implications of these results for a $\delta$-function
excluded volume potential are then discussed. In section~4, the
Brownian dynamics simulation algorithm used in this work is
described. Section~5 is devoted to the development of the Gaussian
approximation for the configurational distribution function. Exact
expressions for linear viscoelastic properties are derived through a
codeformational memory-integral expansion.  In section~6, a first
order perturbation expansion in the strength of the excluded volume
interaction is carried out. This proves to be very helpful in
understanding the nature of the Gaussian approximation.  The results
of the various exact and approximate treatments are compared and
discussed in section~7, and the main conclusions of the paper are
summarized in section~8.

\section{Basic Equations}

The instantaneous configuration of a linear bead-spring chain, which
consists of $N$ beads connected together by $(N-1)$ Hookean springs,
is specified by the bead position vectors ${\br}_{\nu}, \, \nu = 1,2,
\ldots, N,$ in a laboratory-fixed coordinate system. The Newtonian
solvent, in which the chain is suspended, is assumed to have a
homogeneous velocity field---that is, at position $\br$ and time $t$, the 
velocity is given by $\bv (\br, t) = \bv_0 + \bk (t) \cdot {\br}$, 
where $\bv_0$ is a constant vector and $\bk (t)$ is a traceless tensor.

The microscopic picture of the intra-molecular forces within the 
bead-spring chain is one in which the presence of excluded volume 
interactions between the beads causes the chain to swell, while 
on the other hand, the entropic retractive force of the springs 
draws the beads together and opposes the excluded volume force.
This is modeled by writing the potential energy $\phi$ of the 
bead-spring chain as a sum of the potential energy of the springs $S$, and 
the potential energy due to excluded volume interactions $E$. 
The potential energy $S$ is the sum of the potential energies of all
the springs in the chain, and is given by,
\begin{equation}
S = {1 \over 2} \, H \, \sum_{i=1}^{N-1} \, {\bQ}_i \cdot {\bQ}_i
\end{equation}
where, $H$ is the spring constant, and ${\bQ}_i= {\br}_{i+1} -
{\br}_{i}$, is the {\it bead connector} vector between the beads $i$
and $i+1$.  The excluded volume potential energy $E$ is found by
summing the interaction energy over all pairs of beads $\mu$ and
$\nu$,
\begin{equation}
E = {1 \over 2} \sum_{\mu,\nu = 1 
\atop \mu \ne \nu}^N \, E \left( {\br}_{\nu}
- {\br}_{\mu} \right)
\end{equation}
where, $E \left( {\br}_{\nu} - {\br}_{\mu} \right)$ is a short-range
function. It is usually assumed to be a $\delta$-function
potential in static theories for polymer solutions,
\begin{equation}
 E \left( {\br}_{\nu} - {\br}_{\mu} \right) = v \, k_{\rm B} T \, 
\delta \left( {\br}_{\nu} - {\br}_{\mu} \right)
\label{deltapot}
\end{equation}
where, $v$ is the excluded volume parameter (with dimensions of
volume), $ k_{\rm B}$ is Boltzmann's constant, and $T$ is the absolute
temperature.  In this work, we regularise the $\delta$-function
potential, and assume that $E \left( {\br}_{\nu} - {\br}_{\mu}
\right)$ is given by a narrow Gaussian potential,
\begin{equation}
 E \left( {\br}_{\nu} - {\br}_{\mu} \right) = {v \, k_{\rm B} T \over
[2 \pi {\tilde d}^2]^{3 \over 2}} \, \exp \left( - {1 \over 2} 
{ \br_{\nu \mu}^2 
\over {\tilde d}^2} \right)
\label{evpot}
\end{equation}
where, ${\tilde d}$ is a parameter that describes the width of the
potential (it represents, in some sense, the {\em extent} of excluded
volume interactions), and $\br_{\nu \mu} = \br_{\nu}-\br_{\mu}$, is
the vector between beads $\mu$ and $\nu$.  In the limit ${\tilde d}$
tending to zero, the narrow Gaussian potential becomes a
$\delta$-function potential.

The {\em intra-molecular} force on a bead $\nu$, 
${\bF}_{\nu}^{(\phi)} = -  ({\partial \phi / \partial {\br}_{\nu}})$, 
can consequently be written as,
${\bF}_{\nu}^{(\phi)} = {\bF}_{\nu}^{(S)} + {\bF}_{\nu}^{(E)}$, where,
\begin{eqnarray}
{\bF}_{\nu}^{(S)} &=& -  {\partial S \over \partial {\br}_{\nu}} 
= - H \sum_{k=1}^{N-1} {\overline B}_{k \nu} 
\, \bQ_k \label{sprforce} \\
\nonumber \\
{\bF}_{\nu}^{(E)} &=& -  {\partial E \over \partial {\br}_{\nu}} 
= - \sum_{\mu = 1 \atop \mu \ne \nu}^N \, {\partial 
\over \partial \br_\nu} \, E \left( {\br}_{\nu} - {\br}_{\mu} \right) 
\label{evforce}
\end{eqnarray}
In eq~\ref{sprforce}, ${\overline B}_{k \nu}$ 
is an $(N-1) \times N$ matrix defined by, 
${\overline B_{k \nu}} = \delta_{k+1, \; \nu} - \delta_{k \nu}$, 
with $\delta_{k \nu}$ denoting the Kronecker delta. 

For homogeneous flows, the internal configurations of the bead-spring
chain are expected to be independent of the location of the centre of
mass. Consequently, it is assumed that the configurational
distribution function $\psi$ depends only on the $(N-1)$ bead
connector vectors $\bQ_k$. The diffusion equation that governs $ \psi
\, ({\bQ}_1, \ldots, {\bQ}_{N-1}, t)$, for a system with an
intra-molecular potential energy $\phi$ as described above, can then
be shown to be given by,
\begin{eqnarray}
{\partial \, \psi \over \partial t} = &-& \sum_{j=1}^{N-1} \,
{\partial   \over \partial {\bQ}_j} \cdot \biggl( 
\bk \cdot {\bQ}_{j} 
- {H \over \zeta} \, \sum_{k=1}^{N-1} \, A_{jk} \, {\bQ}_k
+ {1 \over \zeta} \, \sum_{\nu=1}^{N}\, {\overline B}_{j \nu} \bF_{\nu}^{(E)}
\, \biggr) \, \psi \nonumber \\
\nonumber \\
&+& {k_{\rm B} T \over \zeta} \, \sum_{j,\, k=1}^{N-1} \, A_{jk} \; 
{\partial \over \partial {\bQ}_j} \cdot
{\partial \psi \over \partial {\bQ}_k} 
\label{diff}
\end{eqnarray}
where, $\zeta$ is the bead friction coefficient (which, for spherical
beads with radius $a$, in a solvent with viscosity $\eta_s$, is given
by the Stokes expression: $\zeta=6 \pi \eta_s a$), and $A_{jk}$ is the
Rouse matrix,
\begin{equation}
A_{jk}=\sum_{\nu=1}^{N}\, {\overline B}_{j \nu} {\overline B}_{k \nu} 
=\cases{ 2&for $\vert {j-k} \vert = 0 $,\cr
\noalign{\vskip3pt}
-1& for $\vert {j-k} \vert =1 $,\cr
\noalign{\vskip3pt}
0 & otherwise \cr} 
\label{rmatrix}
\end{equation}

The time evolution of the average of any arbitrary quantity, carried
out with the configurational distribution function $\psi$, can be
obtained from the diffusion equation. In particular, by multiplying eq
\ref{diff} by ${\bQ}_j {\bQ}_k $, and integrating over all
configurations, the following time evolution equation for the second
moments of the bead connector vectors is obtained,
\begin{eqnarray}
{d \over dt} \, \avel {\bQ}_j  {\bQ}_k \aver &=&
\bk \cdot \avel {\bQ}_j  {\bQ}_k \aver + 
\avel {\bQ}_j  {\bQ}_k \aver \cdot \bk^T + 
{ 2 k_{\rm B} T \over \zeta} \, A_{jk} \, \bu 
\nonumber \\ 
\nonumber \\ 
&-& { H \over \zeta} \, \sum_{m=1}^{N-1} \, \left\lbrace \, 
\avel {\bQ}_j \bQ_m \aver \, A_{mk} + A_{jm} \, \avel {\bQ}_m \bQ_k 
\aver \, \right\rbrace + \bY_{jk}
\label{secmom}
\end{eqnarray}
where, $\bu$ is the unit tensor, and, 
\begin{equation}
\bY_{jk} = { 1 \over \zeta} \, \sum_{\mu =1}^N \, \left\lbrace \,
\avel {\bQ}_j  {\bF}_{\mu}^{(E)} \aver \, {\overline B}_{k \mu} 
+ {\overline B}_{j \mu}  \, \avel {\bF}_{\mu}^{(E)}  
\bQ_k \aver \, \right\rbrace
\label{Yeqn}
\end{equation}

The term $\bY_{jk}$, which arises due to the presence of excluded
volume interactions, does not appear in the second moment equation for
the Rouse model. Due to this term, which in general involves higher
order moments, eq~\ref{secmom}, is not a closed equation for $\avel
{\bQ}_j {\bQ}_k \aver$.  As will be discussed in greater detail in the
section on the Gaussian approximation, finding an approximate solution
for the present model involves making eq~\ref{secmom} a closed
equation for the second moments.

The polymer contribution to the stress tensor---for models with
arbitrary intra-molecular potential forces but no internal
constraints---is given by the {\it Kramers} expression~\cite{bird2}, 
\begin{equation}
\btau^p = - n_{\rm p} H \,  \sum_{k=1}^{N-1} \, 
\avel {\bQ}_k \bQ_k \aver 
+ \bZ + (N-1) \,  n_{\rm p}  k_{\rm B} T \, \bu 
\label{kram}
\end{equation}
where, 
\begin{equation}
\bZ =  n_{\rm p} \, \sum_{\nu=1}^N 
\sum_{k=1}^{N-1} \, B_{\nu k} \, 
\avel \bQ_k {\bF}_{\nu}^{(E)} \aver 
\label{iso}
\end{equation}
Here, $n_{\rm p}$ is the number density of polymers, and $B_{\nu k}$
is a $N \times (N-1)$ matrix defined by, $ B_{\nu k} = k/N - \Theta \,
(k-\nu)$, with $\Theta \, (k-\nu)$ denoting a Heaviside step function.

It is clear from eq~\ref{kram} that there are two reasons why the
presence of excluded volume interactions leads to a stress tensor that
is different from that obtained in the Rouse model.  Firstly, there is
an additional term represented by $\bZ$ which is the {\em direct}
influence of excluded volume effects.  Secondly, there is an {\em
indirect} influence due to a change in the contribution of the term
$\sum_{k=1}^{N-1} \, \avel {\bQ}_k {\bQ}_k \aver$, relative to its
contribution in the Rouse case. For a $\delta$-function excluded
volume potential, it can be shown that the direct contribution to the
stress tensor is isotropic~\cite{doi}. On the other hand, for the
narrow Gaussian potential, $\bZ$ is {\em not} isotropic unless $\tilde
d$ is equal to zero. It is therefore important to use the complete
form of the Kramers expression, eq~\ref{kram}, when carrying out
simulations with an excluded volume potential that is not a
$\delta$-function potential.

All the rheological properties of interest 
can be obtained once the stress tensor in eq~\ref{kram} 
is evaluated. In the next section, a retarded motion expansion 
for the stress tensor is derived. 
 
\section{Retarded Motion Expansion} 
 
A retarded motion expansion for the stress tensor can be obtained 
by extending the derivation carried out previously for the dumbbell 
model~\cite{prakottev} to the case of bead-spring chains. The dumbbell 
model derivation was, in turn, an adaptation of a similar development 
for the FENE dumbbell model~\cite{bird2}. 
The argument in all these cases rests basically 
on seeking a solution of the diffusion
equation, eq~\ref{diff}, of the following form, 
\begin{equation}  
\psi ({\bQ}_1, \ldots, {\bQ}_{N-1}, t) = \psi_{\rm eq} 
({\bQ}_1, \ldots, {\bQ}_{N-1}) \, 
\phi_{\rm fl} \, ({\bQ}_1, \ldots, {\bQ}_{N-1}, t)  
\label{prod}  
\end{equation}  
where, $\psi_{\rm eq}$ is the equilibrium distribution function  
given by, 
\begin{equation} 
\psi_{\rm eq} ({\bQ}_1, \ldots, {\bQ}_{N-1}) = {\cal N_{\rm eq}} 
\, e^{- {\phi / k_{\rm B} T}} 
\label{eqdist} 
\end{equation} 
with ${\cal N_{\rm eq}}$ denoting the normalization constant, and 
$\phi_{\rm fl}$ is the correction to $\psi_{\rm eq}$ due to 
flow---appropriately termed the flow contribution. 

The governing partial differential equation for 
$\phi_{\rm fl} \, ({\bQ}_1, \ldots, {\bQ}_{N-1},t)$  
can be obtained by substituting eq~\ref{prod} into the 
diffusion equation, eq~\ref{diff}. It turns out that, regardless of 
the form of the excluded volume potential, at steady state,   
an exact solution to this partial differential equation can be found  
for all homogeneous {\it potential} flows. For more  
general homogeneous flows, however, one can only obtain a perturbative 
solution of the form, 
\begin{equation}  
\phi_{\rm fl} \, ({\bQ}_1, \ldots, {\bQ}_{N-1},t) =   
1 + \phi_1 + \phi_2 + \phi_3 + \ldots  
\label{flowdistexp}  
\end{equation}  
where $\phi_k$ is of order $k$ in the velocity gradient. 
 
Partial differential equations governing each of the $\phi_k$ may be 
derived by substituting eq~\ref{flowdistexp} into 
the partial differential equation for  $\phi_{\rm fl}$ 
and equating terms of like order.  
The forms of the functions $\phi_k$ can then be guessed by 
requiring that they fulfill certain conditions~\cite{bird2}. In the present instance, 
we only find the form of $\phi_1$, since our interest is confined to 
zero shear rate properties. 
One can show that, 
\begin{equation}  
\phi_{1} =  {\zeta \over 4 k_{\rm B} T } \, \sum_{m, n =1}^{N-1} \, 
C_{m n} \, \bQ_m \cdot {\dot \bgam} \cdot \bQ_n  
\label{phi1} 
\end{equation}  
where, ${\dot \bgam}$ is the rate of strain tensor, 
${\dot \bgam} = \nabla \bv + \nabla 
\bv^T$, and $C_{m n}$ is the Kramers matrix. The Kramers matrix 
is the inverse of the Rouse matrix, and is defined by, 
\begin{equation}  
C_{m n} = \sum_{\nu=1}^{N}\, B_{\nu m} B_{\nu n} = \min \, (m,n) - m n / N
\label{kmatrix} 
\end{equation}

In order to proceed further, we need to show that the present 
model satisfies the Giesekus expression for the stress 
tensor~\cite{bird2}. Upon multiplying eq~\ref{diff} with 
$\sum_{m, n =1}^{N-1} C_{m n} \, {\bQ}_m  {\bQ}_n $, and 
integrating over all configurations, we can show that,
\begin{eqnarray}
{d \over dt} \, \sum_{m, n =1}^{N-1}  
\avel  C_{m n} \, {\bQ}_m  {\bQ}_n \aver -
\sum_{m, n =1}^{N-1}  C_{m n} \, 
\left[ \bk \cdot \avel {\bQ}_m  {\bQ}_n \aver  
+ \avel {\bQ}_m  {\bQ}_n \aver \cdot \bk^T \right] 
\nonumber \\ 
\nonumber \\ 
= { 2 k_{\rm B} T \over \zeta} \, (N-1) \, \bu 
- {2 H \over \zeta} \, \sum_{m=1}^{N-1} \, \avel {\bQ}_m \bQ_m \aver 
+ {2  \over \zeta} \, \sum_{\nu=1}^N \sum_{m=1}^{N-1} \, B_{\nu m} \, 
\avel \bQ_m {\bF}_{\nu}^{(E)} \aver 
\label{gmom}
\end{eqnarray}
On combining this equation with eq~\ref{kram} for the stress tensor, 
it is straight forward to see that the Giesekus expression is 
indeed satisfied. At steady state the Giesekus expression reduces to, 
\begin{equation} 
\btau^p = - { n_{\rm p} \, \zeta \over 2} \sum_{m, n =1}^{N-1} 
C_{m n} \left\lbrace \, \bk \cdot \avel {\bQ}_m  {\bQ}_n \aver  
+ \avel {\bQ}_m  {\bQ}_n \aver \cdot \bk^T \right\rbrace 
\label{gstress} 
\end{equation}
Clearly, the stress tensor at steady state can be found once the
average $\avel {\bQ}_m {\bQ}_n \aver $ is evaluated.  This can be
done, correct to first order in velocity gradients, by using the power
series expansion for $\phi_{\rm fl}$, eq~\ref{flowdistexp}, with the
specific form for $\phi_{1}$ in eq~\ref{phi1}.  The following retarded
motion expansion for the stress tensor, correct to second order in
velocity gradients and valid for arbitrary homogeneous flows, is then
obtained,
\begin{eqnarray} 
\btau^p &=& - { n_{\rm p} \, \zeta \over 2} \sum_{m, n =1}^{N-1} 
 C_{m n} \, \biggl[ \bk \cdot \avel {\bQ}_m  {\bQ}_n \aver_{\rm eq}  
+ \avel {\bQ}_m  {\bQ}_n \aver_{\rm eq}  
\cdot \bk^T \biggr] \nonumber \\
\nonumber \\
&-& { n_{\rm p} \, \zeta^2 \over 8  k_{\rm B} T } 
\sum_{m, n =1}^{N-1} \, \sum_{j, k =1}^{N-1} C_{m n} C_{j k} 
\biggl[ \bk \cdot \avel  {\bQ}_m {\bQ}_n 
({\bQ}_j \cdot  {\dot \bgam} \cdot  {\bQ}_k) \aver_{\rm eq} \nonumber \\
\nonumber \\
&+& \avel  ({\bQ}_j \cdot  {\dot \bgam} \cdot  {\bQ}_k) 
{\bQ}_m  {\bQ}_n \aver_{\rm eq}  \cdot \bk^T \biggr] + \ldots 
\label{retstress} 
\end{eqnarray}
where, $\avel X \aver_{\rm eq}$ denotes the average of any arbitrary 
quantity $X$ with the equilibrium distribution function $\psi_{\rm eq}$. 

One can see clearly from eq~\ref{retstress} that
rheological properties, at small values of the velocity gradient, 
can be obtained by merely evaluating equilibrium averages. The special
case of steady simple shear flow in the limit of zero 
shear rate is considered below. 

\subsection{Zero Shear Rate Viscometric Functions}  

Steady simple shear flows are described by a tensor $\bk$ which 
has the following matrix representation in the laboratory-fixed 
coordinate system, 
\begin{equation} 
\bk={\dot \gamma} \, 
\pmatrix{ 0 & 1 & 0 \cr 
0 & 0 & 0 \cr 
0 & 0 & 0 \cr } 
\label{ssf1} 
\end{equation} 
where ${\dot \gamma}$ is the constant shear rate. 
 
The three independent material functions used to characterize such 
flows are the viscosity, $\eta_p$, and the first and second normal 
stress difference coefficients, $\Psi_1 \,\,{\rm and}\,\, \Psi_2$, 
respectively.  These functions are defined by the following relations 
\cite{bird1}, \begin{equation} \tau_{xy}^p = - {\dot \gamma}\, \eta_p 
\, ; \quad \quad \tau_{xx}^p- \tau_{yy}^p = - {\dot \gamma^2}\, \Psi_1 
\, ; \quad \quad \tau_{yy}^p- \tau_{zz}^p = - {\dot \gamma^2}\, \Psi_2 
\label{sfvis} \end{equation} 

The components of the stress tensor in simple shear flow, 
for small values of  the shear rate ${\dot \gamma}$,  
can be found by substituting eq~\ref{ssf1} for the rate  
of strain tensor, into eq~\ref{retstress}. This leads to, 
\begin{eqnarray} 
\tau_{xy}^p &=& - { n_{\rm p} \, \zeta  \, {\dot \gamma} 
\over 2} \sum_{m, n =1}^{N-1} C_{m n} \, 
\avel Y_m  Y_n \aver_{\rm eq}  
- { n_{\rm p} \, \zeta^2  \, {\dot \gamma}^2 \over 4  k_{\rm B} T } 
\sum_{m, n =1}^{N-1} \, \sum_{p, q =1}^{N-1} C_{m n} C_{p q} 
\avel  Y_m Y_n X_p Y_q \aver_{\rm eq} \nonumber \\
\nonumber \\
\tau_{xx}^p &=& - n_{\rm p} \, \zeta \, {\dot \gamma} 
\sum_{m, n =1}^{N-1} C_{m n} \, \avel X_m  Y_n \aver_{\rm eq}  
- { n_{\rm p} \, \zeta^2  \, {\dot \gamma}^2 \over 2  k_{\rm B} T } 
\sum_{m, n =1}^{N-1} \, \sum_{p, q =1}^{N-1} C_{m n} C_{p q} 
\avel  X_m Y_n X_p Y_q \aver_{\rm eq}\nonumber \\
\nonumber \\
\tau_{yy}^p &=& \tau_{zz}^p = 0 
\label{comp} 
\end{eqnarray}
where, $(X_m, Y_m, Z_m)$ are the Cartesian components of the bead 
connector vector $\bQ_m$. 

Using the symmetry property of the potential energy 
$\phi$, which remains unchanged when the sign of the $Y_k$ component
of all the bead connector vectors $\bQ_k ; k=1,2, \ldots, (N-1)$, 
is reversed, we can show that,
$\avel X_m  Y_n \aver_{\rm eq} =0$. 
From the definitions of the viscometric functions in 
eq~\ref{sfvis}, it is straight forward to show that, 
in the limit of zero shear rate, the following exact expressions 
for the zero shear rate viscometric functions are obtained. 
\begin{eqnarray} 
\eta_{p,0} &=&  {n_{\rm p} \, \zeta \over 6} \,
\sum_{m, n =1}^{N-1} C_{m n} \, \avel \bQ_m \cdot \bQ_n \aver_{\rm eq}  
\label{etap0} \\
\nonumber \\ 
\Psi_{1,0} &=&  { n_{\rm p} \, \zeta^2  \over 2  k_{\rm B} T } \, 
\sum_{m, n =1}^{N-1} \, \sum_{p, q =1}^{N-1} C_{m n} C_{p q} 
\, \avel  X_m Y_n X_p Y_q \aver_{\rm eq}
\label{Psi10} \\
\nonumber \\
\Psi_{2,0} &=& 0
\label{Psi20}  
\end{eqnarray}  
In order to derive eq~\ref{etap0}, we have used the fact that, since
$\phi$ is the same function of $X_p$, $Y_p$, and $Z_p$ for all $p$,
$\avel X_p X_q \aver_{\rm eq} = \avel Y_p Y_q \aver_{\rm eq} = \avel
Z_p Z_q \aver_{\rm eq}$. Equation~\ref{Psi20} indicates that the
inclusion of excluded volume interactions alone is not sufficient to
lead to the prediction of a non-zero second normal stress
difference. The proper inclusion of hydrodynamic interaction is
required.

It is interesting to note, by making use of eq~\ref{etap0}, that the
mean square radius of gyration at equilibrium, which is defined
as~\cite{bird2},
\begin{equation} 
\avel R_g^2 \, \aver_{\rm eq} = { 1\over N} \sum_{\nu =1}^{N} 
\int \! \! d{\bQ}_1 d{\bQ}_2 \ldots d{\bQ}_{N-1} \, (\br_{\nu} -\br_c) \cdot 
(\br_{\nu} -\br_c) \, \psi_{\rm eq} 
\label{radg} 
\end{equation}
(where, $\br_c$ is the position of the center of mass),
is related to the zero shear rate viscosity by, 
\begin{equation} 
\eta_{p,0} = { n_{\rm p} \, \zeta  \over 6} \, N \, \avel R_g^2 \, \aver_{\rm eq}
\label{etaradg} 
\end{equation}

An alternative expression for the zero shear rate viscosity, 
which will prove very useful subsequently, can also be obtained from 
eq~\ref{etap0}, 
\begin{equation} 
\eta_{p,0} = {n_{\rm p} \, \zeta \over 12 N} \,
\sum_{\nu, \mu =1}^{N} \, \avel \br_{\nu \mu}^2  \aver_{\rm eq} 
\label{etapmunu} 
\end{equation}
In order to derive eqs~\ref{etaradg} and~\ref{etapmunu}, 
equations which relate the bead connector vector coordinates 
to bead position vector coordinates, summarized for example, in Chapter~11 
and Table~15.1-1 of Chapter~15 of the text book by Bird et al.~\cite{bird2},  
have been used. 

The evaluation of the equilibrium averages in eqs~\ref{Psi10} 
and~\ref{etapmunu}, for various values of the parameters in the 
narrow Gaussian potential, and for various chain lengths $N$, 
have been carried out here with the help of Brownian dynamics simulations.
More details of these simulations are given subsequently. 
In the special case of the extent of excluded volume interactions  
$\tilde d$ going to zero or infinity,
we had shown earlier for a Hookean dumbbell model that the values of 
$\eta_{p,0}$ and $\Psi_{1,0}$ remain unchanged from the values that  
they have in the absence of excluded volume interactions~\cite{prakottev}.
In the next section, we consider the same limits for the more general case of 
bead-spring chains of arbitrary (but finite) length. 

\subsection{The Limits $ {\tilde d} \to 0$ and ${\tilde d} \to \infty$} 

The average in eq~\ref{etapmunu} can be evaluated with the 
distribution function 
$\psi_{\rm eq} ({\bQ}_1, \ldots, {\bQ}_{N-1})$, 
or equivalently, with the distribution function 
$P_{\rm eq} (\br_{\nu \mu})$, which is a contracted distribution function 
for each vector $\br_{\nu \mu}$, and which is defined by, 
\begin{equation} 
P_{\rm eq} (\br_{\nu \mu}) = \int \! \! d{\bQ}_1 d{\bQ}_2 \ldots d{\bQ}_{N-1} \, 
\delta \left( \br_{\nu \mu} - \sum_{j = \mu}^{\nu - 1 } 
\bQ_j \right) \psi_{\rm eq}  
\label{rmunudist1} 
\end{equation}
We have assumed here, without loss of generality, that $\nu > \mu$. 

In the Rouse model, as is well known, the equilibrium distribution 
function is Gaussian, 
\begin{equation}
\psi_{\rm eq}^R \, (\,{\bQ}_1, \ldots, \, {\bQ}_{N-1} \,)
= \prod_{j=1}^{N-1} \, \Biggl( \, {H \over 2  \pi  k_{\rm B} T } \, \Biggr)^{3/2} \,
\exp \, \Biggl( - {H \over 2 \,k_{\rm B} T } \; {\bQ}_j \cdot {\bQ}_j \Biggr)   
\label{rousequidis}
\end{equation}
A superscript or subscript `$R$' on any quantity will henceforth 
indicate a quantity defined or evaluated in the Rouse model. The 
distribution function $P_{\rm eq}^R (\br_{\nu \mu})$ can then be evaluated, 
by substituting eq~\ref{rousequidis}
and the Fourier representation of a $\delta$-function,  into 
eq~\ref{rmunudist1}~\cite{bird2}, 
\begin{equation} 
P_{\rm eq}^R \, (\br_{\nu \mu}) = 
\Biggl( \, {H \over 2  \pi  k_{\rm B} T \, | \nu -\mu | } \, \Biggr)^{3/2} \,
\exp \, \Biggl( - { H \over 2 \, |\nu - \mu| \, k_{\rm B} T } \; \br_{\nu \mu}^2 \Biggr)   
\label{rmunudist2} 
\end{equation}
The absolute value $|\nu - \mu|$ indicates that this expression is valid 
regardless of whether $\nu$ or $\mu$ is greater. 
This is another well known result of the Rouse model, namely, 
at equilibrium, the vector 
$\br_{\nu \mu}$ between any two beads $\mu$ and $\nu$, also 
obeys a Gaussian distribution. 

A similar procedure can be adopted to evaluate 
$P_{\rm eq} (\br_{\nu \mu})$, in the presence of excluded 
volume interactions, by substituting eq~\ref{eqdist}
and the Fourier representation of a $\delta$-function,  into 
eq~\ref{rmunudist1}. We show in appendix~\ref{peq} that,
\begin{equation} 
\lim_{{\tilde d} \to 0 \atop {\rm or}, \, {\tilde d} \to \infty} 
P_{\rm eq} (\br_{\nu \mu}) = P_{\rm eq}^R \, (\br_{\nu \mu}) 
\label{rmunudistlim} 
\end{equation}
As a result, for all quantities $X(\br_{\nu \mu})$, such that 
the product $X(\br_{\nu \mu}) \, P_{\rm eq} (\br_{\nu \mu})$
remains bounded for all $\br_{\nu \mu}$, 
$\lim_{{\tilde d} \to 0 \atop {\rm or}, \, {\tilde d} \to \infty}  
\avel X(\br_{\nu \mu}) \aver_{\rm eq} = \avel X(\br_{\nu \mu}) \aver_{\rm eq}^R$. 
It follows from eq~\ref{etapmunu} that, 
\begin{equation} 
\lim_{{\tilde d} \to 0 \atop {\rm or}, \, {\tilde d} \to \infty}
\eta_{p,0} = \eta_{p,0}^R 
\label{etaplim} 
\end{equation}
Thus, the polymer contribution to the viscosities in the 
limit of zero shear rate, for chains of arbitrary (but 
finite) length, in (i) the presence of $\delta$-function 
excluded volume interactions, and (ii) the absence of excluded volume 
interactions (the Rouse model), 
are identical to each other. Brownian dynamics simulations, 
details of which are given in the section below, indicate that this is
also true for the first normal stress difference coefficients. 

\section{Brownian Dynamics Simulations} 

The equilibrium averages in eqs~\ref{Psi10} 
and~\ref{etapmunu}, as mentioned above, can be evaluated 
with the help of Brownian dynamics simulations. As a result, 
exact numerical predictions of the zero shear rate 
viscometric functions can be obtained. 
Brownian dynamics simulations basically involve the numerical 
solution of the Ito stochastic differential equation that 
corresponds to the diffusion equation, eq~\ref{diff}. 
Using standard methods~\cite{ottbk} to transcribe a Fokker-Planck equation  
to a stochastic differential equation, one can show that 
eq~\ref{diff} is equivalent to the following system 
of $(N-1)$ stochastic differential equations for the 
connector vectors $\bQ_j$,
\begin{equation} 
d \bQ_j = \left\lbrace  \bk \cdot \bQ_j - {1 \over \zeta} \sum_{k=1}^{N-1} 
A_{jk} \, {\partial \phi \over \partial \bQ_k } \right\rbrace \, dt 
+ \sum_{\nu=1}^N  \sqrt{ {2 k_{\rm B} T \over \zeta} } \, 
{\overline B}_{j \nu} \; d \bW_{\nu}  
\label{ito} 
\end{equation} 
where, $\bW_{\nu}$ is a $3 N$ dimensional Wiener process.  

A second order predictor-corrector algorithm with time-step 
extrapolation~\cite{ottbk} was used for the numerical solution of 
eq~\ref{ito}. Steady-state expectations at equilibrium 
were obtained by setting $\bk=0$, and simulating a single long trajectory. 
This is justified based on the assumption of ergodicity~\cite{ottbk}. 

\section{The Gaussian Approximation} 

A crucial step in the calculation of the rheological properties
predicted by the present model is the evaluation of the complex
moments that occur in Kramers expression. The Gaussian
approximation---which has previously been shown to be useful in the
treatment of hydrodynamic interaction and internal viscosity
effects~\cite{prakbk,schiebiv,wedgeiv}---consists essentially of
reducing complex higher order moments to functions of only second
order moments by assuming that the non-equilibrium configurational
distribution function is a Gaussian distribution, and subsequently,
evaluating these second order moments by integrating a time evolution
equation.

For the narrow Gaussian potential, the complex moment $\avel {\bQ}_k
{\bF}_{\mu}^{(E)} \aver$, which appears in the quantity $\bZ$ on the
right hand side of Kramers expression, eq~\ref{kram}, can be rewritten
in terms of averages of the form: $\avel {\bQ}_k {\bQ}_n E \left(
{\br}_{\nu} - {\br}_{\mu} \right) \aver$. Assuming that $\psi$ is a 
Gaussian distribution of the form, 
\begin{equation}
\psi \, ({\bQ}_1, \ldots, {\bQ}_{N-1}, t) \, = \, {\cal N} (t) \,
\exp \big[ - {1 \over 2} \sum_{j, \, k} \, {\bQ}_j \cdot 
({\bs}^{- 1})_{jk}
\cdot {\bQ}_k \big]
\label{gauss}
\end{equation}
where, the $(N-1) \times (N-1)$ matrix of tensor components $\bsjk$
(with $\bsjk= \avel {\bQ}_j{\bQ}_k \aver$ and $ \bsjk=\bs_{kj}^T $) 
uniquely characterizes the Gaussian distribution and ${\cal N}(t)$ is
the normalization factor, and using general decomposition
rules for the moments of a Gaussian distribution~\cite{ottga}, one can
show that,
\begin{eqnarray}
\avel {\bQ}_m  {\bQ}_n E \left( {\br}_{\nu} 
- {\br}_{\mu} \right) \aver = 
{v \, k_{\rm B} T \over \left( 2 \pi \right)^{3/ 2} } \, 
{1 \over \sqrt{ \det \left( [ {\tilde d}^2 \, \bu + \avel
{\br}_{\nu \mu} {\br}_{\nu \mu} \aver ] \right) } } \times 
\nonumber \\
\nonumber \\
\left\lbrace \avel {\bQ}_m  {\bQ}_n \aver - 
\avel {\bQ}_m {\br}_{\nu \mu} \aver 
\cdot \left[ {\tilde d}^2 \, \bu + \avel {\br}_{\nu \mu} {\br}_{\nu \mu} 
\aver \right]^{-1} \cdot \avel {\br}_{\nu \mu}  {\bQ}_n \aver \right\rbrace
\label{decomp}
\end{eqnarray}
The vector ${\br}_{\nu \mu}$ also obeys a Gaussian distribution
since it is a sum of Gaussian distributed bead connector vectors. As a
result, the right hand side of eq~\ref{decomp} involves only second
moments, and averages which can be evaluated by Gaussian integrals.

In the Gaussian approximation therefore, Kramers expression 
for the stress tensor can be rewritten as,
\begin{equation}
\btau^p = - n_{\rm p} H \,  \sum_{k=1}^{N-1} \, \bs_{kk} 
+ \bZ + (N-1) \,  n_{\rm p}  k_{\rm B} T \, \bu 
\label{gausskram}
\end{equation}
where,
\begin{equation}
\bZ =  {1 \over 2} \, z^* \, n_{\rm p}  k_{\rm B} T  
\sum_{\nu,  \, \mu=1 \atop \nu \ne \mu }^N  
\bsh_{\nu \mu} \cdot \bPi(\bsh_{\nu \mu})
\label{gaussiso}
\end{equation}
In eq~\ref{gaussiso}, the function $\bPi(\bsh_{\nu \mu})$ is given by,
\begin{equation}
\bPi(\bsh_{\nu \mu}) = { \left[ {d^*}^2 \, \bu + \bsh_{\nu \mu} \right]^{-1}  
\over 
\sqrt{ \det \left( [ {d^*}^2 \, \bu + \bsh_{\nu \mu} ] \right) }}
\label{Pifun}
\end{equation}
with the tensors $\bsh_{\nu \mu}$ defined by,
\begin{equation}
\bsh_{\nu \mu} = \bsh_{\nu \mu}^T = \bsh_{\mu \nu} = {H \over k_B T }
\,\, \sum_{j,k \,= \,\min(\mu,\nu)}^{ \max(\mu,\nu)-1} \, \bsjk
\end{equation}
The quantities $z^*$ and $d^*$ are non-dimensional versions of the
two parameters, $v$ and $\tilde d$, which characterize the narrow Gaussian
potential. They are defined by,
\begin{equation} 
z^* =  v \, \left( \, {H \over 2 \pi k_B T} \right)^{3 \over 2}; 
\quad d^* = {\tilde d} \, \sqrt{ H \over k_B T}
\label{zandd}
\end{equation}
While $z^*$ measures the strength of the excluded volume interaction, 
$d^*$ is a measure of the extent of excluded volume interaction. 

In the limit of $d^* \to 0$, it is straight forward to see
that the tensor $\bZ$ becomes isotropic. As a result, the direct
contribution to the stress tensor has no influence on the rheological
properties of the polymer solution only when a $\delta$-function potential
is used to represent excluded volume interactions. 

All that remains to be done in order to evaluate the stress tensor is
to find the components of the covariance matrix $\bsjk$. A system of
$9 \, (N-1)^2$ coupled ordinary differential equations for $\bsjk$ can
be obtained from the time evolution equation for the second moments,
eq~\ref{secmom}. As mentioned earlier, in the presence of excluded
volume interactions, eq~\ref{secmom} also involves higher order
moments due to the occurrence of the term $\bY_{jk}$, and 
consequently, it is not in general a closed equation for the second
moments. However, these higher order moments can also be reduced to
second order moments with the help of the decomposition result,
eq~\ref{decomp}. In the Gaussian approximation, the second moment
equation can therefore be rewritten as,
\begin{equation}
{d \over dt} \, \bsjk =
\bk \cdot \bsjk + 
\bsjk  \cdot \bk^T + 
{ 2 k_{\rm B} T \over \zeta} \, A_{jk} \, \bu 
- { H \over \zeta} \, \sum_{m=1}^{N-1}  \left[ 
\bs_{jm} \, A_{mk} + A_{jm} \, \bs_{m k} \right] + \bY_{jk}
\label{gaussecmom}
\end{equation}
where, 
\begin{equation}
\bY_{jk} = {z^*} \, \left( { H \over \zeta} \right) \, \sum_{m=1}^{N-1}  
\left[ \bs_{jm} \cdot \bDel_{km} + \bDel_{jm} \cdot \bs_{m k} \right]
\label{gaussY}
\end{equation}
In eq~\ref{gaussY}, the $(N-1) \times (N-1)$ matrix of tensor
components $\bDel_{jm}$ is defined by, 
\begin{equation}
\bDel_{jm} = \sum_{\mu=1}^{N}  \biggl\lbrace 
( B_{j+1,\, m} -  B_{\mu m}) \, \bPi (\bsh_{j+1,\, \mu})
- ( B_{j m} -  B_{\mu m}) \, \bPi (\bsh_{j \mu}) \biggr\rbrace
\label{Del}
\end{equation}

For any homogeneous flow, rheological properties predicted by the
Gaussian approximation can be obtained by appropriately choosing the
tensor $\bk$, solving the differential equations, eqs
\ref{gaussecmom}, for $\bsjk$, and substituting the result into
Kramers expression, eq~\ref{gausskram}. In this paper, we confine
attention to the prediction of linear viscoelastic properties, namely,
material functions in small amplitude oscillatory shear flow, and zero
shear rate viscometric functions. 

Linear viscoelastic properties predicted by the Gaussian approximation
can be obtained by constructing a codeformational memory-integral
expansion. This is done by expanding the tensors $\bsjk$, in terms of
deviations from their isotropic equilibrium solution, up to first
order in velocity gradient,
\begin{equation}  
\bsjk = f_{jk} \, \bu + \beps_{jk} + \ldots 
\label{linvisexp}  
\end{equation} 
where, the tensors $f_{jk} \, \bu$ represent equilibrium second
moments in the Gaussian approximation, and the tensors $\beps_{jk}$
are the first order corrections. Since the details of the 
calculation are not very illuminating, they are given in
appendix~\ref{codef}, and only the results are summarized below. 

The first order codeformational memory-integral expansion derived by
the above procedure has the form, 
\begin{equation}
\btau^p= -   \int_{- \infty}^t d\!s \, G(t-s)\, 
{\bgam}_{[1]}(t,s) 
\label{memexp}
\end{equation}
where, ${\bgam}_{[1]}$ is the codeformational rate-of-strain
tensor~\cite{bird1}, and the memory function $G(t)$ is given by
eq~\ref{memfun} in appendix~\ref{codef}. This expression can
now be used to obtain exact expressions for material functions in
small amplitude oscillatory shear flow, and for the zero shear rate
viscosity and first normal stress difference coefficient in steady
shear flow, as shown below. 

Small amplitude oscillatory shear flow is characterized by a tensor
$\bk(t)$ given by, 
\begin{equation} 
\bk(t)={\dot \gamma}_0 \, \cos \, \omega t 
\pmatrix{ 0 & 1 & 0 \cr 0 & 0 & 0 \cr 0 & 0 & 0 \cr } \label{usf3} 
\end{equation} 
where, ${\dot \gamma_0}$ is the amplitude, and $\omega$ 
is the frequency of oscillations in the plane of flow.  The $yx$ 
component of the polymer contribution to the shear stress is 
then defined by~\cite{bird1}, 
\begin{equation} 
\tau_{yx}^p=- 
\eta^\prime(\omega)\, {\dot \gamma}_0 \, \cos \, \omega t - 
\eta^{\prime\prime}(\omega)\, {\dot \gamma}_0 \, \sin \, \omega t 
\label{usf4} 
\end{equation} 
where $\eta^\prime$ and $ \eta^{\prime\prime}$ are the material 
functions characterizing oscillatory shear flow. They can be represented 
in a combined form as the complex viscosity, 
$\eta^* =\eta^\prime - i \,\eta^{\prime\prime}$, and $\eta^*$ 
can be found, in terms of the relaxation modulus, from the  
expression 
\begin{equation} 
\eta^*= \int_0^\infty G(s)\, e^{-i \omega s} \, d\!s
\label{comvis}
\end{equation}
Upon substituting eq~\ref{memfun} for the memory function $G(s)$ into
eq~\ref{comvis}, one obtains the predictions of the Gaussian
approximation for $\eta^\prime$ and $\eta^{\prime\prime}$. These are
given by eqs~\ref{etaprime} in appendix~\ref{codef}. 

The zero shear rate viscosity $\eta_{p,0}$ and the zero shear 
rate first normal stress difference coefficient $\Psi_{1,0}$, 
can be obtained from the complex viscosity in 
the limit of vanishing frequency, 
\begin{equation} \eta_{p,0} = 
\lim_{\omega\to 0} \, \eta^{\prime} (\omega) \, ; \quad \quad 
\Psi_{1,0} = \lim_{\omega\to 0} {2 \, \eta^{\prime\prime} (\omega) 
\over \omega} 
\label{usf8} 
\end{equation} 
The predictions of the zero shear rate viscometric functions by the 
Gaussian approximation are given by eqs~\ref{stetap0}
and~\ref{stpsi10} in appendix~\ref{codef}. They are compared 
with the exact results, eqs~\ref{etap0} and~\ref{Psi10},
evaluated by Brownian dynamics simulations, in section~7 below. 

\section{First Order Perturbation Expansion}

The retarded motion expansion, eq~\ref{retstress}, which was obtained
by carrying out a perturbation expansion of the distribution function
$\psi$, in terms of velocity gradients, is valid for arbitrary
strength of the excluded volume interaction.  In this section, using
arguments similar to those in the papers by \"Ottinger and
co-workers~\cite{ottrabrg,zylkarg,ottrg}, we derive a perturbation
expansion of $\btau^p$ in the strength of excluded volume interaction,
which is valid for arbitrary shear rates. A significant benefit of the
perturbation expansion will be a better understanding of the nature of
the Gaussian approximation.

The distribution function $\psi$ may be written as $\psi_R +
\psi_{z^*}$, where $\psi_R$ is the distribution function in the
absence of excluded volume, i.e. in the Rouse model, and $\psi_{z^*}$
is the correction to first order in the strength of the excluded
volume interaction. Since $\psi_R$ is Gaussian, it has the form given
by eq~\ref{gauss}, with ${\cal N}(t)$ replaced by ${\cal N}_R(t)$, and
$\bsjk$ replaced by $\bs^R_{jk} =\avel \bQ_j \bQ_k \aver_R $.  The
second moments $\avel \bQ_j \bQ_k \aver$ can then be expanded to first
order as, $\avel \bQ_j \bQ_k \aver = \bs^R_{jk} + \avel \bQ_j \bQ_k
\aver_{z^*}$.

On substituting this expansion into eq~\ref{secmom}, and equating
terms of like order, the second moment equation can be separated into
two equations, a zeroth-order equation and a first-order equation. The
zeroth-order equation, which is the second moment equation of the
Rouse model, is linear in $\bs^R_{jk}$, and has the following explicit
solution,
\begin{equation}
\bs^R_{jk} =  {k_{\rm B} T \over H} \left\lbrace 
\delta_{jk} \, \bu + \, \int_{- \infty}^{t} \! \! \! d  \! s \,  
\, {\cal E} \left[ - {2 H \over \zeta} (t-s) A \right]_{jk} 
\, {\bgam}_{[1]}(t, s) \right\rbrace
\label{qqtheta}
\end{equation}
where, $\cal E$ is an exponential operator. Properties of exponential
operators that operate on $(N-1)^2 \times (N-1)^2$ matrices are
discussed in appendix~\ref{codef}. The exponential operators used in
this section have similar properties, but operate on $(N-1) \times
(N-1)$ matrices.

The first-order second moment equation has the form,
\begin{eqnarray}
{d \over dt} \, \avel {\bQ}_j {\bQ}_k \aver_{z^*} &=& \bk \cdot \avel
{\bQ}_j {\bQ}_k \aver_{z^*} + \avel {\bQ}_j {\bQ}_k \aver_{z^*} \cdot
\bk^T \nonumber \\ \nonumber \\ &-& \left( { H \over \zeta} \right) \,
\sum_{m=1}^{N-1} \, \left\lbrace \, \avel {\bQ}_j \bQ_m \aver_{z^*} \,
A_{mk} + A_{jm} \, \avel {\bQ}_m \bQ_k \aver_{z^*} \, \right\rbrace +
\bY^R_{jk}
\label{fordsecmomz}
\end{eqnarray}
where, $\bY^R_{jk}$ is given by eq~\ref{Yeqn}, with the averages on
the right hand side evaluated with $\psi_R$, i.e., $\avel \cdots
\aver$ are replaced with $\avel \cdots \aver_R$. Since $\psi_R$ is a
Gaussian distribution, the decomposition result, eq~\ref{decomp}, with
$\avel \cdots \aver$ replaced with $\avel \cdots \aver_R$, can be used
to reduce $\bY^R_{jk}$ to a function of second moments alone. This
leads to, 
\begin{equation}
\bY^R_{jk} = z^* \, \left( { H \over \zeta} \right) \, \sum_{m=1}^{N-1}  
\left[ \bs^R_{jm} \cdot \bDel^R_{km} + \bDel^R_{jm} \cdot \bs^R_{m k} \right]
\label{fordY}
\end{equation}
In eq~\ref{fordY}, $\bDel^R_{jm}$ is given by eq~\ref{Del}, with $\bsjk$
replaced by $\bs^R_{jk}$ in the definition of $\bshmn$ on the right
hand side.  Equation~\ref{fordsecmomz} is a system of linear
inhomogeneous ordinary differential equations, whose solution is,
\begin{eqnarray}
\avel \bQ_j \bQ_k \aver_{z^*}  &=&  \sum_{r,s =1}^{N-1} 
\int_{- \infty}^{t} \!\! d\!s \,  
{\cal E} \left[ - {H \over \zeta} (t-s) A \right]_{jr} 
 \, \bE(t, s) \cdot \bY^R_{rs} (s) \cdot  \bE{^T}(t, s) \nonumber \\
\nonumber \\
&\times& {\cal E} \left[ - {H \over \zeta} (t-s) A \right]_{sk} 
\label{qqz}
\end{eqnarray}
where, $\bE$ is the displacement gradient tensor~\cite{bird1}. 

It is immediately clear from eq~\ref{fordsecmomz} that the Gaussian
approximation is exact to first order in the strength of excluded
volume interaction. This follows from the fact that it could have also
been derived by expanding eq~\ref{gaussecmom} to first order in
$z^*$. It will be seen later that this property of the Gaussian
approximation, is helpful in elucidating its nature.

The first order perturbation expansion for the stress tensor can be
obtained by expanding Kramers expression, eq~\ref{kram}, to first
order in $z^*$. After reducing complex moments evaluated with the
Rouse distribution function to second moments, the stress tensor can
be shown to depend only on second moments through the relation,
\begin{equation} 
\btau^p = - n_{\rm p} H \,  \sum_{k=1}^{N-1} \, \bs^R_{kk} 
- n_{\rm p} H \,  \sum_{k=1}^{N-1} \, \avel \bQ_k \bQ_k \aver_{z^*} 
+ \bZ^R + (N-1) \,  n_{\rm p}  k_{\rm B} T \, \bu 
\label{fordkram}
\end{equation}
where, $\bZ^R$ is given by eq~\ref{gaussiso}, with $\bsjk$ replaced by
$\bs^R_{jk}$ in the definition of $\bshmn$ on the right hand side.
Equations~\ref{qqtheta} and~\ref{qqz} may then be used to derive the
following first order perturbation expansion for the stress tensor in
arbitrary homogeneous flows,
\begin{eqnarray}
\btau^p &=& - n_{\rm p} k_BT  \sum_{r,s =1}^{N-1} 
\int_{- \infty}^{t} \!\! d\!s \,  
{\cal E} \left[ - {2 H \over \zeta} (t-s) A \right]_{sr} 
\bE(t, s) \cdot \Biggl\lbrace  \left( \bk(s) 
+\bk^T(s) \right) \, \delta_{r s} \nonumber \\ 
\nonumber \\ 
&+& \left( {H \over k_{\rm B} T} \right) \, \bY^R_{rs}(s) \Biggr\rbrace
\cdot  \bE{^T}(t, s) + \bZ^R + (N-1) \,  n_{\rm p}  k_{\rm B} T \, \bu 
\label{pertau}
\end{eqnarray}
Note that $\bZ^R$, the direct contribution to the stress tensor, is
isotropic only in the limit $d^* \to 0$. We now consider the special
case of steady shear flow, and obtain the zero shear rate viscometric
functions.

\subsection{Steady Shear Flow}

In order to obtain zero shear rate viscometric functions correct to
first order in $z^*$, it is necessary to evaluate the time integrals
in eqs~\ref{qqtheta} and \ref{pertau}, and to evaluate the quantities
$\bY^R_{jk}$ and $\bZ^R$ in steady shear flow. The results of these
calculations are given below, while the details are given in
appendix~\ref{visco}.

The {\em excluded volume contributions} to the zero shear rate
viscometric functions (correct to first order in $z^*$) obtained
by setting $\dot \gamma$ equal to zero in eqs~\ref{pertetap} to
\ref{pertpsi2} of appendix~\ref{visco} are, 
\begin{eqnarray} 
\eta_{p,0}^{(E)} &=&  {1 \over 2}  \, \lambda_H^2 \, z^*  
\sum_{\mu, \nu =1 \atop \mu \ne \nu}^{N}
{1 \over \left( {d^*}^2 + S_{\mu \nu}^{(0)} \right)^{7/2} }
\, \biggl[ \, S_{\mu \nu}^{(0)} \, S_{\mu \nu}^{(1)}  
+ {d^*}^2 \, S_{\mu \nu}^{(1)} \, \biggr]
\label{pertetap0} \\ 
\nonumber \\
\Psi_{1,0}^{(E)} &=&   \lambda_H^2 \, z^*  
\sum_{\mu, \nu =1 \atop \mu \ne \nu}^{N}
{1 \over \left( {d^*}^2 + S_{\mu \nu}^{(0)} \right)^{7/2}} 
\, \left[ \, 2 \, S_{\mu \nu}^{(2)} 
\left( {d^*}^2 + S_{\mu \nu}^{(0)} \right)
- S_{\mu \nu}^{(1)} \, S_{\mu \nu}^{(1)} \right]
\label{pertpsi10} \\
\nonumber \\
\Psi_{2,0}^{(E)} &=&  0
\label{pertpsi20}  
\end{eqnarray}  
where, the time constant $\lambda_H = (\zeta / 4H)$ has been
introduced previously in appendix~\ref{codef}, and the quantities
$S_{\mu \nu}^{(m)}$, which occur in these functions and which were
introduced earlier by \"Ottinger~\cite{ottrg}, are defined by,
\begin{equation}
S_{\mu \nu}^{(m)} = 2^m \, 
\sum_{j,k \,= \,\min(\mu,\nu)}^{ \max(\mu,\nu)-1} \,
\, C^m_{jk}
\label{smunu}
\end{equation}

The first order perturbation theory predictions of the zero shear rate
viscometric functions given above are compared with exact Brownian
dynamics simulations and the Gaussian approximation in section~7. We
first, however, examine the role of the parameters $d^*$ and $z^*$ in the
present model, by considering the end-to-end vector at equilibrium in
the limit of large $N$.

\subsection{The Equilibrium End-to-End Vector For Large Values of $N$}

The second moment of the end-to-end vector 
$\br$ at equilibrium is given by the expression,
\begin{equation} 
\avel \br \br \aver_{\rm eq} =  \sum_{j, k=1}^{N-1}  
\avel \bQ_j \bQ_k \aver_{\rm eq}
\label{endtoend1}
\end{equation}
For the Rouse model, $\bs^R_{jk} \, \bigg\vert_{\rm eq} = \left(
k_{\rm B} T / H \right) \, \delta_{jk} \, \bu$.  One can show, from
eq~\ref{qqz}, that the first order correction to the second moments
has the following form at equilibrium,
\begin{equation}
\avel \bQ_j \bQ_k \aver_{z^*}  \Bigg\vert_{\rm eq} 
=  \left( {\zeta \over H} \right)
\sum_{r,s =1}^{N-1} R^{-1}_{jk, \, r s} \, \bY^R_{rs} 
\end{equation}
where, the $(N-1)^2 \times (N-1)^2$ matrix $R_{jk, \, m n }$ is defined by,
$R_{jk, \, mn} =  A_{jm} \, \delta_{kn} + \delta_{jm}  \, A_{kn}$,
and $\bY^R_{jk}$ has the form,
\begin{equation} 
\bY^R_{jk} \Bigg\vert_{\rm eq} =  z^*  
\, {k_{\rm B} T \over 2 \zeta } \sum_{\mu, \nu =1 \atop \mu \ne \nu}^{N}
{1 \over \left( {d^*}^2 + S_{\mu \nu}^{(0)} \right)^{5/2}} 
\left( \sum_{m, n =1 }^{N-1} \theta(\mu,m,n,\nu) \, R_{jk,\, mn} \right)  \bu 
\label{bY}
\end{equation}
Note that the function $\theta(\mu,m,n,\nu)$ has been introduced
previously in appendix~\ref{codef} (see eq~\ref{theta}).  It follows
that the mean square end-to-end vector at equilibrium, correct to
first order in $z^*$, is given by,
\begin{equation} 
\avel \br^2 \aver_{\rm eq} =  {3 \, k_{\rm B} T \over H} \left[
(N-1)  + {1 \over 2} \, 
z^*   \sum_{\mu, \nu =1 \atop \mu \ne \nu}^{N}
{ | \mu - \nu |^2 \over \left( {d^*}^2 
+  | \mu - \nu | \right)^{5/2}} 
\right]
\label{endtoend2}
\end{equation}

We now consider the limit of a large number of beads, $N$. In this limit, 
the sums in eq~\ref{endtoend2} can be replaced by integrals.
Introducing the following variables,
\begin{equation} 
x = {\mu \over N}; \quad  y = {\nu \over N}; \quad  d = {d^* \over \sqrt{N}} 
\label{ddef}
\end{equation} 
and exploiting the symmetry in $x$ and $y$, we obtain,
\begin{equation} 
\avel \br^2 \aver_{\rm eq} =  {3 \, k_{\rm B} T \over H} \, N \left\lbrace 
1  + z^* \, \sqrt{N}  \mathop{ \int_{0}^{1} \! \! d x
\int_{0}^{x} \! \! d y}_{x > y + c } \, 
{  (x -y)^2 \over \left( d^2 +  x - y \right)^{5/2}} 
\right\rbrace
\label{endtoend3}
\end{equation}
where, $c$ is a {\em cutoff} parameter of order $1/N$ which 
accounts for the fact that $\mu \ne \nu$. 

It is clear from  eq~\ref{endtoend3} that the excluded volume 
corrections to the Rouse end-to-end vector are proportional 
to $z^* \, \sqrt{N}$. As a result, the proper perturbation 
parameter to choose is $z \equiv z^* \, \sqrt{N}$, and not $z^*$. 
This is a very well known result of the theory of polymer 
solutions~\cite{doi,declos,schaf}, and indicates that 
a perturbation expansion in  $z^* $ becomes useless for long chains. 

The integrals in eq~\ref{endtoend3} can be performed analytically.
However, we are interested only in the form of eq~\ref{endtoend3},
which leads to a very valuable insight.  Defining the quantity
$\alpha$, which is frequently used to represent the {\em swelling} of
the polymer chain at equilibrium due to excluded volume effects,
\begin{equation} 
\alpha^2 = {\avel \br^2 \aver_{\rm eq} \over \avel \br^2 \aver_{\rm eq}^R }  
\label{swell}
\end{equation}
we can see that in the limit of long chains, $\alpha = \alpha \, ( z,
d \, )$.  In other words, $\alpha$ depends asymptotically only on the
parameters $z$ and $d$, and not on the chain length $N$. We shall see
later that this insight is very useful in understanding the results of
Brownian dynamics simulations, and the Gaussian approximation.

\begin{figure}[!htb] \centerline{ \epsfxsize=4in \epsfbox{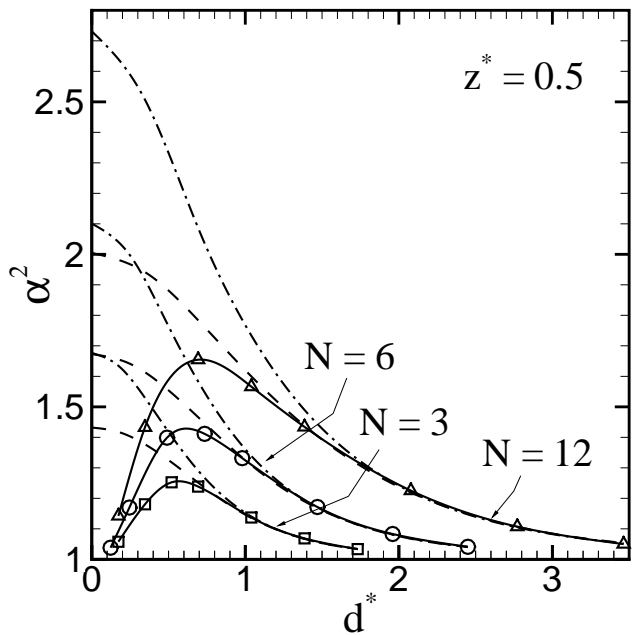}}  
\caption{   Equilibrium swelling of the end-to-end vector versus
the extent of excluded volume interaction $d^*$, at a constant value
of the strength of the interaction $z^*$, for three different values
of chain length, $N$. The non-dimensional parameters $z^*$ and $d^*$
characterize the narrow Gaussian potential, and are defined in
eq~\ref{zandd}. The squares, circles and triangles are results of
Brownian dynamics simulations, the dashed and dot-dashed lines are the
approximate predictions of the Gaussian approximation, and the first
order perturbation theory, respectively. The error bars in the
Brownian dynamics simulations are smaller than the size of the
symbols, and the continuous curves through the symbols are drawn for
guiding the eye.}  
\end{figure}

\begin{figure}[!htb] \centerline{ \epsfxsize=4in \epsfbox{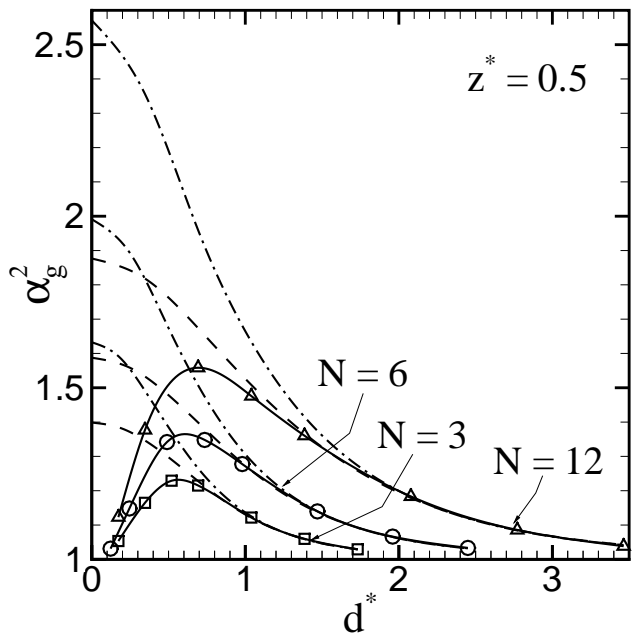}}
\caption{   Swelling of the radius of gyration versus $d^*$, at a
constant value of $z^*$, for three different values of $N$. Note that
$\alpha^2_g = {\eta_{p,0} / \eta_{p,0}^R}$. The symbols are as
indicated in the caption to Figure~1. The error bars in the Brownian
dynamics simulations are smaller than the size of the symbols, and the
continuous curves through the symbols are drawn for guiding the eye.}
\end{figure}

\begin{figure}[!htb] \centerline{ \epsfxsize=4in \epsfbox{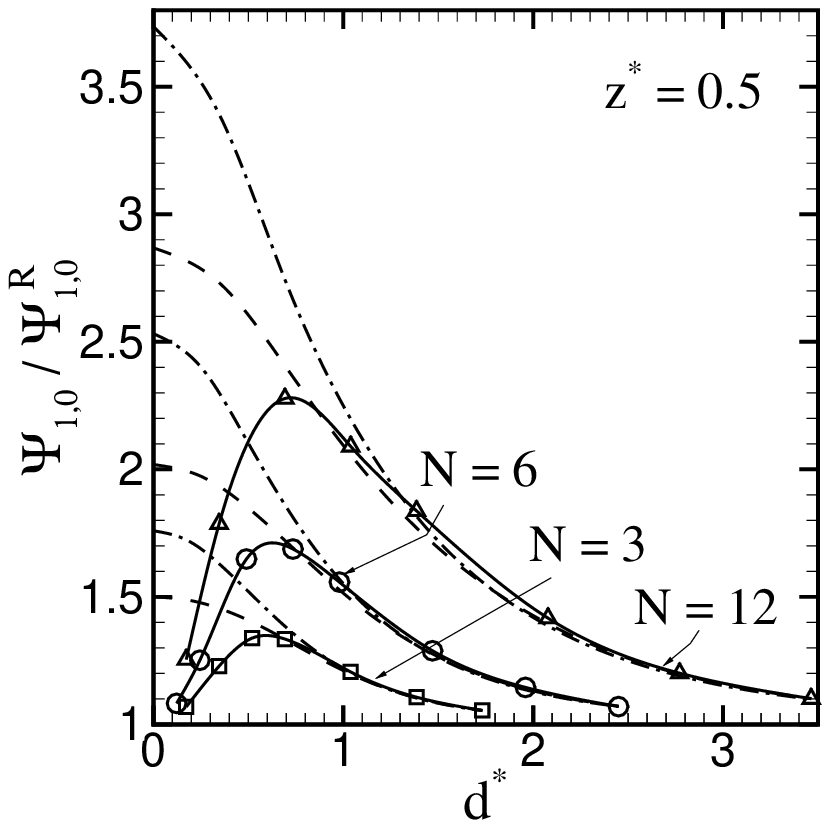}}
\caption{   Ratio of the zero shear rate first normal stress
difference coefficient in the presence of excluded volume interactions
to the zero shear rate first normal stress difference coefficient in
the Rouse model versus $d^*$, at a constant value of $z^*$, for three
different values of $N$. The symbols are as indicated in the caption
to Figure~1. The error bars in the Brownian dynamics simulations are
smaller than the size of the symbols, and the continuous curves
through the symbols are drawn for guiding the eye.}
\end{figure}

\begin{figure}[!htb] \centerline{ \epsfxsize=4in \epsfbox{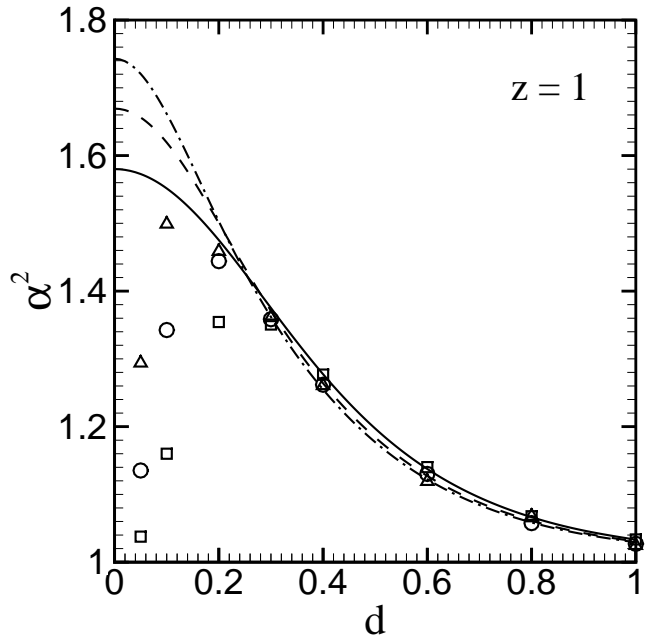}}
\caption{   Equilibrium swelling of the end-to-end vector versus
the re-scaled extent of excluded volume interaction $d = d^* / \sqrt
N$, at a constant value of the re-scaled strength of the interaction
$z= {z^*} \sqrt N$, for three different values of $N$. The squares,
circles and triangles represent the results of Brownian dynamics
simulations for $N$ equal to 6, 12 and 24 beads, respectively. The
continuous, dashed, and dot-dashed curves are the approximate
predictions of the Gaussian approximation for $N$ equal to 6, 12 and
24 beads, respectively. The error bars in the Brownian dynamics
simulations are smaller than the size of the symbols.}
\end{figure}

\begin{figure}[!htb] \centerline{ \epsfxsize=4in \epsfbox{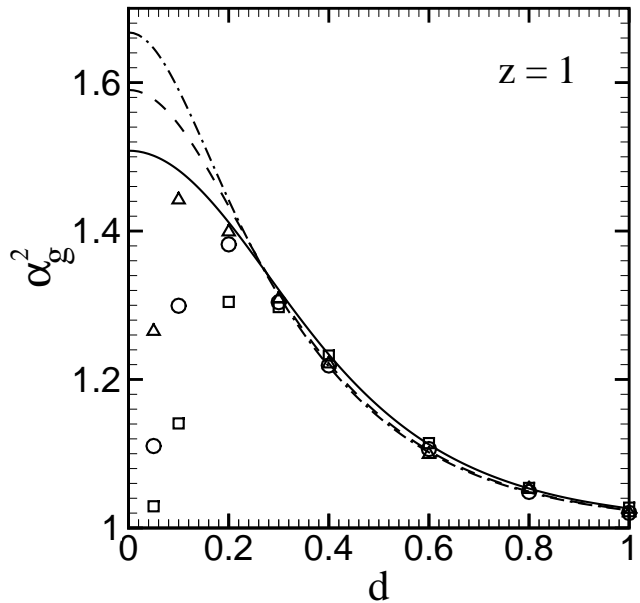}}
\caption{   Swelling of the radius of gyration versus $d$, at a
constant value of $z$, for three different values of $N$. The symbols
are as indicated in the caption to Figure~4. The error bars in the
Brownian dynamics simulations are smaller than the size of the
symbols.}
\end{figure}

\begin{figure}[!htb] \centerline{ \epsfxsize=4in \epsfbox{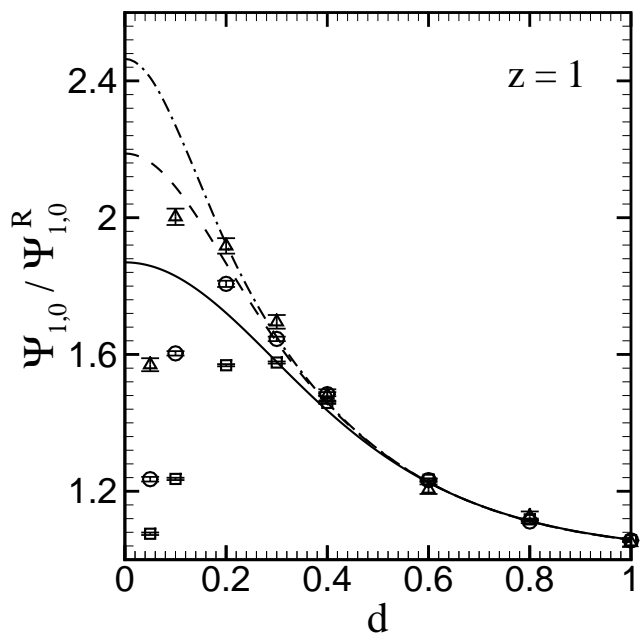}}
\caption{   Ratio of the zero shear rate first normal stress
difference coefficient in the presence of excluded volume interactions
to the zero shear rate first normal stress difference coefficient in
the Rouse model versus $d$, at a constant value of $z$, for three
different values of $N$. The symbols are as indicated in the caption
to Figure~4.}
\end{figure}

\begin{figure}[!htb] \centerline{ \epsfxsize=4in \epsfbox{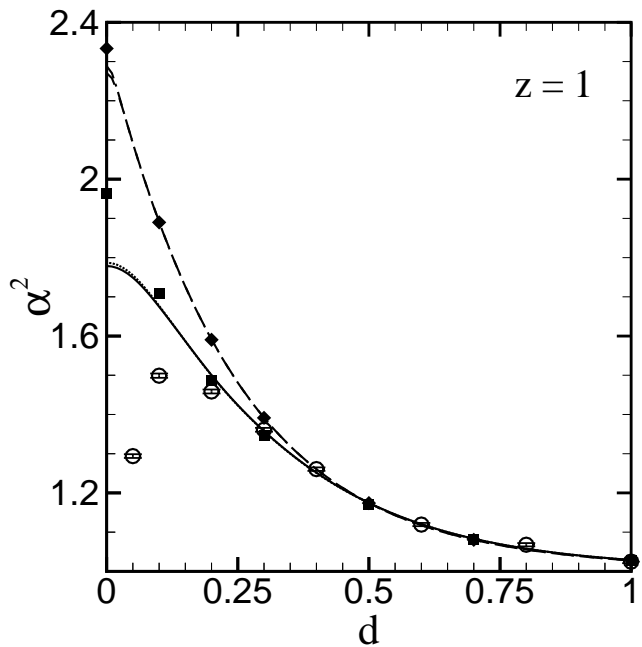}}
\caption{   Equilibrium swelling of the end-to-end vector versus
$d$, at a constant value of $z$, for different $N$.  The continuous
and dotted curves are the predictions of the Gaussian approximation
for $N$ equal to 36 and 40 beads, respectively. The filled squares are
the asymptotic predictions of the Gaussian approximation, obtained by
numerical extrapolation of finite chain data to the limit of infinite
chain length. The dashed and dot-dashed curves are the predictions of
the first order perturbation theory for $N$ equal to 500 and 1000
beads, respectively. The filled diamonds are the predictions of the
first order perturbation theory in the long chain limit, obtained by
carrying out the integrals in eq~\ref{endtoend3} analytically. The
circles, with error bars, are the results of Brownian dynamics
simulations for $N=24$.}
\end{figure}

\begin{figure}[!htb] \centerline{ \epsfxsize=4in \epsfbox{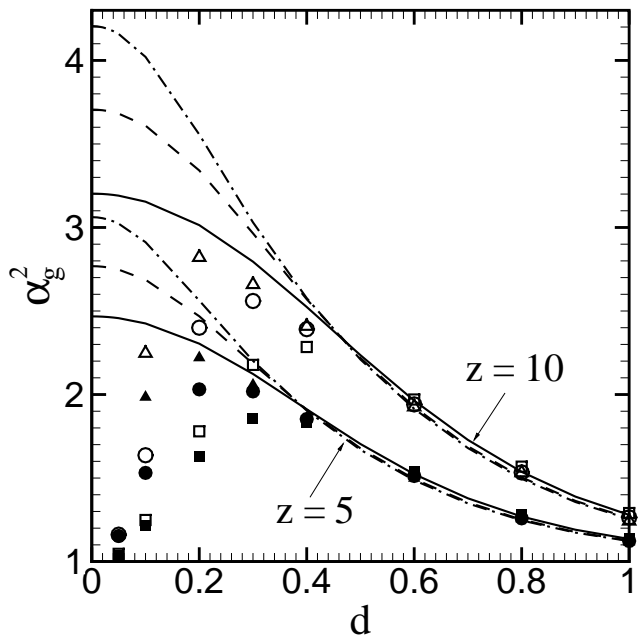}}
\caption{   Swelling of the radius of gyration versus $d$, at two
values of $z$, for three different values of $N$. The squares, circles
and triangles (filled for $z=5$ and empty for $z=10$) represent the
results of Brownian dynamics simulations for $N$ equal to 6, 12 and 24
beads, respectively. The continuous, dashed, and dot-dashed
curves are the approximate predictions of the Gaussian approximation
for $N$ equal to 6, 12 and 24 beads, respectively. The error bars in
the Brownian dynamics simulations are smaller than the size of the
symbols.}
\end{figure}

\begin{figure}[!htb] \centerline{ \epsfxsize=4in \epsfbox{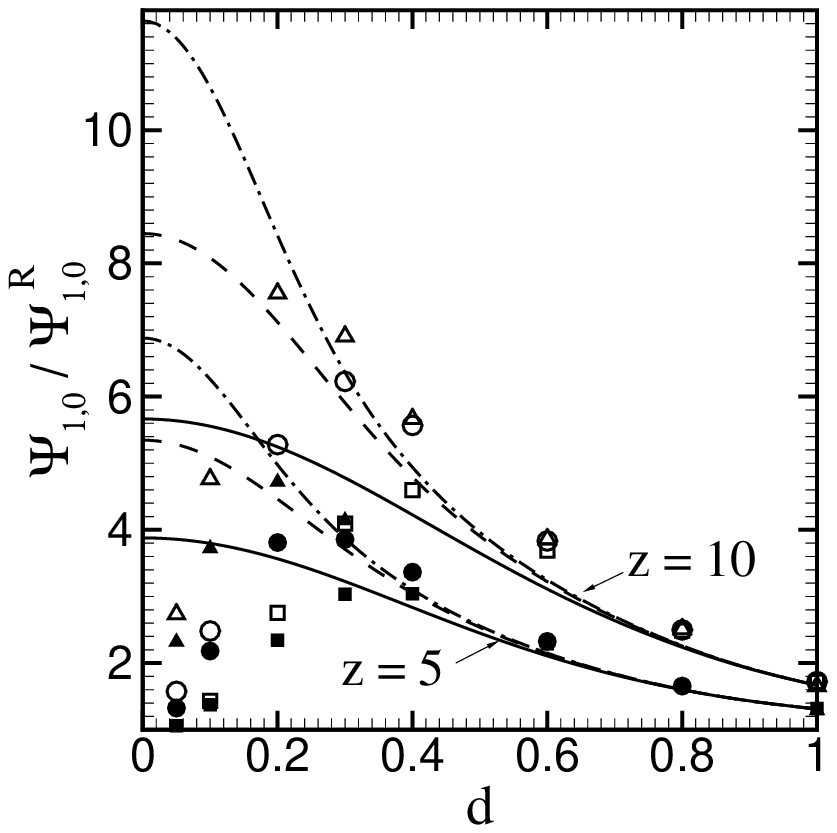}}
\caption{   Ratio of the zero shear rate first normal stress
difference coefficient in the presence of excluded volume interactions
to the zero shear rate first normal stress difference coefficient in
the Rouse model versus $d$, at two values of $z$, for three different
values of $N$. The symbols are as indicated in the caption to
Figure~8. The error bars in the Brownian dynamics simulations are
smaller than the size of the symbols.}
\end{figure}

\section{Equilibrium Swelling and Zero Shear Rate \\ Viscometric Functions}

The prediction of equilibrium properties and zero shear rate 
viscometric functions, by Brownian dynamics simulations, 
the Gaussian approximation and the first order perturbation 
expansion, are compared in this section.  
Before doing so, it is appropriate to note that an equilibrium 
property, frequently defined in static theories of polymer 
solutions, namely, the swelling of the radius of gyration, 
$\alpha^2_g$, can be found from the following expression,
\begin{equation}
\alpha^2_g = {\avel R_g^2 \, \aver_{\rm eq} \over 
\avel R_g^2 \, \aver_{\rm eq}^R} = {\eta_{p,0} \over \eta_{p,0}^R}
\label{alphag}
\end{equation}
because of the relation between the radius of gyration 
and the zero shear rate viscosity, eq~\ref{etaradg}.
Plots of $\alpha^2_g$ in this section must, therefore, 
also be seen as plots of the ratio of the zero shear rate 
viscosity in the presence of excluded volume interactions to 
the zero shear rate viscosity in the Rouse model.

Figures~1 to~3 are plots of $\alpha^2$,
$\alpha^2_g$ and $(\Psi_{1,0}/ \Psi_{1,0}^R) $ versus $d^*$,
respectively, at a constant value of $z^*= 0.5$, for three different
chain lengths, $N=3$, $N=6$ and $N=12$.  The squares, circles and
triangles are exact results of Brownian dynamics simulations for the
narrow Gaussian potential, the dashed lines are the predictions of the
Gaussian approximation, and the dot-dashed curves are the predictions
of the first order perturbation theory.

In the limit $d^* \to 0$ and for large values of $d^*$, for 
all the values of chain length $N$, the Brownian dynamics simulations 
reveal that equilibrium and zero shear rate properties tend to Rouse 
model values. In the case of $\alpha^2$ and $\alpha^2_g$, 
this is expected because of the rigorous result, eq~\ref{rmunudistlim}.
An immediate implication of this behavior is that, 
for chains of arbitrary but finite length, 
it is not fruitful to use a $\delta$-function 
potential to represent excluded volume interactions. 
On the other hand, the figures seem to suggest
that a finite range of excluded volume interaction is required to cause
an increase from Rouse model values. Both the first order perturbation 
theory and the Gaussian approximation predict a significant change 
from Rouse model values in the limit $d^* \to 0$. In the case of a
dumbbell model, we were able to rigorously understand the origin of 
these spurious results~\cite{prakottev}. The incorrect term-by-term 
integration of a series that was not uniformly convergent was found
to be the source of the problem. Since first order perturbation 
theory is the basis for renormalisation group calculations, the invalidity 
of the $\delta$-function potential, which is frequently used in 
these calculations, is at first sight worrisome. 
However, we shall see below that the use of a $\delta$-function potential 
may be legitimate when both the limits, $N \to \infty$ 
and $d^* \to 0$, are considered.

Figures~1 to~3 show that there exists a 
threshold value of $d^*$ at which, the results of the Gaussian 
approximation and the first order perturbation 
theory, first agree with exact Brownian dynamics simulations. 
This is consistent with the first order perturbation 
theory predictions of the end-to-end vector, eq~\ref{endtoend2}, 
and the viscometric functions, eqs~\ref{pertetap0} 
and~\ref{pertpsi10}, which reveal that, excluded volume 
corrections to the Rouse model decrease with increasing values 
of $d^*$. The threshold value of $d^*$ at which the approximations 
become accurate, increases as 
$N$ increases. This is a consequence of the well known result, 
which was demonstrated in section~6, that excluded volume corrections 
scale as ${z^*} \, \sqrt N$. Note, however, that 
the Gaussian approximation always becomes accurate 
at a smaller threshold value of $d^*$ than the first order perturbation 
theory. The Gaussian approximation, while being a  
non-perturbative approximation, is nevertheless, exact to first order 
in $z^*$. Consequently, as mentioned earlier, it might be 
considered to consist of an infinite number of higher order terms, 
and can be expected to be more accurate than the results of the first order 
perturbation theory. 

All the results in Figures~1 to~3 are entirely 
consistent with the results obtained earlier with a dumbbell model for 
the polymer molecule. However, in the case of the dumbbell model, 
the dependence of the quality of the approximations on the chain 
length $N$, could not be examined. The results in Figures~1 
to~3 seem to suggest that the Gaussian approximation has a 
rather limited validity, since for a given value of $z^*$ and $d^*$, 
it gets progressively worse as the chain length $N$ increases. This is 
in fact not a realistic picture---as is revealed below---when the data 
is reinterpreted in terms of a different set of coordinates. 

Figures~4 to~6 are plots of $\alpha^2$, $\alpha^2_g$ and 
$(\Psi_{1,0}/ \Psi_{1,0}^R) $ versus $d= (d^* / \sqrt N)$, respectively, 
at a constant value of $z= {z^*} \sqrt N = 1.0$, 
for three different chain lengths, $N=6$, $N=12$ and $N=24$. 
Before we discuss the figures, it is appropriate to make a few remarks 
about the choice of the variables in terms of which the data are displayed. 
Firstly, we choose $z$ as the measure of the strength of excluded volume 
interaction because perturbation theory clearly reveals that
excluded volume corrections scale as 
${z^*} \, \sqrt N$. A constant value 
of $z$, as $N$ increases, implies that ${z^*}$ must 
simultaneously decrease in order to keep the relative 
role of excluded volume interactions the same. Secondly, we 
choose the $x$-axis coordinate as $d= (d^* / \sqrt N)$, because, 
as was shown in section~6, perturbation theory in the limit of long chains 
indicates that when the data is displayed in terms of $d$ and $z$, 
all the curves should collapse on to a single curve as $N \to \infty$. 
The parameter $d$ may be considered to be the extent 
of excluded volume interaction, measured as a fraction of the 
unperturbed (i.e., Rouse) root mean square end-to-end vector 
$\sqrt{\avel \br^2 \, \aver_{\rm eq}^R}$. 
 
We first discuss the results of exact Brownian dynamics simulations 
displayed in Figures~4 to~6. As in 
Figures~1 to~3, all the properties 
start at Rouse values at $d=0$, go through a maximum as $d$
increases, and then finally decrease once more towards Rouse values 
with the continued increase of $d$. However, as the chain 
length increases, the values seem to rise increasingly 
more rapidly from the Rouse values at $d=0$, towards the 
maximum value. In other words, the slope at the origin seems 
to be getting steeper as $N$ increases. Indeed, the trend of the 
data leads one to speculate that, in the limit $N \to \infty$, 
the data might be singular at $d=0$, and consequently legitimize, 
in this limit, the use of a $\delta$-function excluded volume potential. This 
conclusion is of course only speculative, and needs to be established more 
rigorously. It has not been possible to examine more closely, with the help of 
Brownian dynamics simulations, the behavior at small values of $d$ 
for larger values of $N$, because of the excessive CPU time 
that is required. In terms of the non-dimensional time 
$t^* = (t/\lambda_H)$, for $N = 24$, a  run for two 
non-dimensional time steps $\Delta t^* = 0.1$ and $\Delta t^* = 0.08$, 
required roughly 65 hours of CPU time on a SGI Origin2000 with a 195 MHz 
processor. 

When viewed in terms of $z$ and $d$, the Gaussian approximation is 
revealed to be far more satisfactory than appeared at first sight in 
Figures~1 to~3. Indeed, for relatively small
values of $d$, where the Gaussian approximation is inaccurate at small 
values of chain length, the Gaussian approximation seems to be becoming
more accurate as $N$ increases. One might expect that as $N \to \infty$, 
the Gaussian approximation becomes
accurate for an increasingly larger range of $d$ values. However, 
as will perhaps become clearer with the results discussed below, 
it appears that, for a given value of $z$, there exists a 
threshold value of $d$, {\em below} which the Gaussian approximation 
will be {\em inaccurate}, no matter how large a choice of $N$ is made.
The reason for this behavior is related to a feature that is just 
noticeable in these figures---curves for different values of 
$N$ appear to be converging to an asymptote. This feature will become 
much clearer in Figure~7, and will be discussed in greater 
detail below. 

For the sake of clarity, the predictions of the first order perturbation 
theory are not displayed in Figures~4 to~6. In 
contrast to the situation in Figures~1 to~3, 
where the accuracy of the first order perturbation theory becomes
progressively worse as $N$ increases, its accuracy appears {\em frozen} 
when viewed in terms of $z$ and $d$. In other words, for 
different---sufficiently large---values 
of $N$, the first order perturbation 
theory first becomes accurate at the same threshold value of $d$.  
As in the case of the predictions of the Gaussian approximation, 
curves for different values of $N$ appear to be converging to a 
common asymptote. This can be seen clearly in Figure~7. 

Figure~7 displays plots of $\alpha^2$ versus $d$, 
for different chain lengths, at a constant value of $z=1$. 
It clearly reveals the fact that, both in the Gaussian approximation 
and in the first order perturbation theory, curves for different 
values of $N$ collapse on to a single curve in the limit $N \to 
\infty$. A similar approach to an asymptotic limit is 
observed as $N \to \infty$, in the predictions of $\alpha^2_g$ 
and $(\Psi_{1,0}/ \Psi_{1,0}^R)$ by both the approximations, when they 
are plotted versus $d$. The results of Brownian dynamics 
simulations for $N=24$ are also plotted in Figure~7. 
They indicate that for $z=1$, asymptotic values have 
already been reached by Brownian dynamics simulations, 
at this relatively small value of $N$, for $d \ge 0.3$.
One expects that as $N$ increases, asymptotic values will be reached 
for smaller and smaller values of $d$. 

The asymptotic values predicted by the first order 
perturbation theory were obtained by carrying out the integrals 
in eq~\ref{endtoend3} analytically. It is worth 
noting that the convergence to the asymptotic value is quite slow 
as $d \to 0$. On the other hand, the asymptotic values predicted 
by the Gaussian approximation were obtained by a numerical 
procedure, as discussed below. 

In the Gaussian approximation, calculation of the equilibrium and zero
shear rate quantities requires the evaluation of the equilibrium
moments $f_{jk}$. These are found here, as mentioned in
appendix~\ref{codef}, by numerical integration of the system of
ordinary differential equations, eq~\ref{equisecmom}, using a simple
Euler scheme, until steady state is reached (the symmetry in $j$ and
$k$ is used to reduce the number of equations by a factor of two).  In
addition, the evaluation of $\eta_{p,0}$ and $\Psi_{1,0}$ requires the
inversion of the $(N-1)^2 \times (N-1)^2$ matrix ${\overline A}_{jk,
\, mn}$ (see eqs~\ref{stetap0} and \ref{stpsi10}). 
As a result, the CPU time scales as $N^6$, and makes the task
of generating data for large values of $N$ extremely computationally
intensive. We have explored the predictions of chains up to a maximum
of $N=40$, since for this value of $N$, a single run on the SGI
Origin2000 computer required approximately 54 hours of CPU time.  The
asymptotic values in Figure~7 were obtained by the
following procedure.  For $z=1$, equilibrium and zero shear rate data,
consisting of property values at different pairs of values $(d,N)$,
were first compiled by performing a large number of runs for various
values of $N$ as a function of $d$.  A specific value of $d$ was then
chosen, and assuming that the various properties were functions of $1
/ \sqrt N$, the values for different $N$ were extrapolated to the
limit $N \to \infty$ using a rational function extrapolation
algorithm~\cite{numrec}. The choice of $1 / \sqrt N$ is motivated by the
fact that the leading correction to the integrals in
eq~\ref{endtoend3} is of order $1 / \sqrt N$~\cite{schaf}.

The quality of Gaussian approximation as a function of the variable
$z$, for the quantities $\alpha^2_g$ and $(\Psi_{1,0}/ \Psi_{1,0}^R)
$, is displayed in Figures~8 and~9,
respectively.  The behavior of $\alpha^2$ has not been displayed as it
is very similar to that of $\alpha^2_g$. It is clear from these
figures that for a given value of $N$, the threshold value of $d$
beyond which the Gaussian approximation is accurate increases as $z$
increases.  On the other hand, as in the case of $z=1$, for a fixed
value of $z$, the accuracy of the Gaussian approximation seems to be
increasing with $N$, for small values of $d$. There is, however,
clearly a limit to this accuracy. As $N$ becomes large, the results of
the exact Brownian dynamics simulations and the Gaussian approximation
approach asymptotic values, and consequently, no further change can be
noticed with changing $N$.  Figures~8 and~9 seem
to indicate that at small values of $d$, while the asymptotic values
of Brownian dynamics simulations lie {\em below} the asymptotic values
of the Gaussian approximation for $\alpha^2_g$, the opposite is true
for $(\Psi_{1,0}/ \Psi_{1,0}^R)$. A clearer picture would be obtained
if it were possible to carry out Brownian dynamics simulations with
larger values of $N$.

Typical experimental values of $z$ lie 
in the range $0 \le z \le 15$~\cite{schaf}. As we have seen above, 
for large enough values of $d$, the Gaussian approximation 
remains accurate for a significant fraction of values of $z$ 
in this range. Since corrections to the Rouse model
due to excluded volume interactions decrease with increasing 
shear rate, we can anticipate that the accuracy of the Gaussian 
approximation will improve as the shear rate increases. Furthermore,
since the Gaussian approximation is extremely accurate for the 
treatment of hydrodynamic interaction effects, and since hydrodynamic 
interaction is likely to be the dominant effect in a combined theory 
of hydrodynamic interaction and excluded volume~\cite{rem1}, it is 
perhaps fair to say that the results obtained so far clearly 
indicate the practical usefulness of the Gaussian approximation. 

\section{Conclusions}

The influence of excluded volume interactions on the linear
viscoelastic properties of a dilute polymer 
solution has been studied with the help
of a narrow Gaussian excluded volume potential that acts between pairs
of beads in a bead-spring chain model for the polymer molecule.  Exact
predictions of the model have been obtained by carrying out Brownian
dynamics simulations, and approximate predictions have been obtained
by two methods---firstly, by carrying out a first order perturbation
expansion in the strength of excluded volume interaction, and
secondly, by introducing a Gaussian approximation for the
configurational distribution function.

The most appropriate way to represent the results of 
model calculations has been shown to be in terms of 
a suitably normalized strength of excluded volume interaction $z$, 
and a suitably normalized extent of excluded volume interaction $d$.
When the results are viewed in terms of these variables, 
the following conclusions can be drawn:
\begin{enumerate} 
\item  The use of a $\delta$-function excluded volume 
potential (which is the narrow Gaussian excluded 
volume potential in the limit $d \to 0$) is not 
fruitful for chains with an arbitrary, but finite, number 
of beads $N$, because it leads to the prediction of 
properties identical to the Rouse model. The narrow 
Gaussian potential with a  finite, non-zero, extent of 
interaction $d$, on the other hand, causes a swelling 
of the polymer chain at equilibrium, and an increase 
in the zero shear rate properties from their
Rouse model values. 
\item Curves for different---but sufficiently large---values 
of chain length $N$, collapse on to a unique asymptotic 
curve in the limit $N \to \infty$. The manner in 
which the  results of Brownian dynamics simulations approach 
the asymptotic behavior indicates that there might be 
a singularity at $d=0$, and consequently, the use of 
a $\delta$-function excluded volume potential might be 
justified in the limit of infinite chain length. 
\item The accuracy of the first order perturbation
expansion becomes independent of $N$ for large $N$. 
For a given value of $z$, there exists a threshold 
value of $d$ beyond which the results of the 
first order perturbation theory agree with the exact 
results of Brownian dynamics simulations. 
\item As in the case of the first order perturbation 
expansion, there exists a threshold 
value of $d$ beyond which the results of the 
Gaussian approximation agree with exact 
results. For a given value of $z$, this threshold
value for accuracy is smaller than the threshold 
value in the first order perturbation theory. 
The accuracy of the Gaussian approximation decreases 
with increasing values of $z$. 

\end{enumerate}

Explicit expressions for the end-to-end vector and the 
viscometric functions in terms of the model parameters, 
obtained by carrying out a first 
order perturbation expansion, enable one to understand the behavior 
of the Gaussian approximation. This is because the Gaussian 
approximation is shown here to be exact to first order in $z$. 

The accuracy of the Gaussian approximation, for a given value of $z$
and $d$, is expected to improve as the shear rate increases.  This
follows from the fact that corrections to the Rouse model, due to
excluded volume interactions, decrease with increasing shear rate.
Viscometric functions at non-zero shear rate predicted by Brownian
dynamics simulations, the Gaussian approximation, and the first order
perturbation expansion, will be compared in a subsequent publication.

The advantage of using the narrow Gaussian potential to represent
excluded volume interactions is that the accuracy of 
approximate solutions can be assessed by comparison with 
exact results. This is in contrast with the situation for approximate 
renormalisation group calculations based on a $\delta$-function 
excluded volume potential, whose accuracy can only be judged 
by comparison with experiment, or with Monte Carlo simulations 
based on a different excluded volume potential. 

The results obtained here indicate that the Gaussian approximation is
an accurate approximation for describing excluded volume interactions,
albeit within a certain range of parameter values. Since the
usefulness of the Gaussian approximation has already been established
for the treatment of hydrodynamic interactions, it is clearly
worthwhile to examine the quality of the Gaussian approximation in a
model for the combined effects of hydrodynamic interaction and
excluded volume.

\vskip15pt

{\bf Acknowledgement.} Support for this work through a 
grant III. 5(5)/98-ET from the Department of Science and Technology, 
India, is acknowledged. A significant part of this work was carried 
out while the author was an Alexander von Humboldt 
fellow at the Department of Mathematics, University of 
Kaiserslautern, Germany. The author would like to thank 
Professor H. C. \"Ottinger for carefully reading the 
manuscript, and for helpful suggestions. Thanks are also due 
to the High Performance Computational Facility at the 
University of Kaiserslautern for providing the use of 
their computers.

\appendix

\section{$P_{\rm eq} (\br_{\nu \mu})$ in the limit ${\tilde d} \to 0$ 
or ${\tilde d} \to \infty$}
\label{peq}
Upon substituting eq~\ref{eqdist} and the Fourier representation 
of a $\delta$-function into eq~\ref{rmunudist1}, and rearranging terms, 
one obtains,
\begin{equation} 
P_{\rm eq} (\br_{\nu \mu}) = {1 \over (2 \pi)^3} \int \! \! d \bfk 
\, e^{i  \, \br_{\nu \mu} \cdot \bfk} \left\lbrace {\cal N_{\rm eq}} 
\int \! \! d{\bQ}_1 \ldots d{\bQ}_{N-1} \, 
e^{-  \left[ \left(\phi / k_{\rm B} T \right)+ i \left( \sum_{j = 
\mu}^{\nu - 1 } \bQ_j \right) \cdot \bfk \right] } \right\rbrace 
\label{rmunudist3} 
\end{equation}
We now consider the integral within braces on the right hand side of 
eq~\ref{rmunudist3}, and take up the integration over the 
bead connector vector $\bQ_1$. Separating out all the terms containing the 
vector $\bQ_1$, we can rewrite this integral as, 
\begin{eqnarray} 
{\cal N_{\rm eq}} \int \! \! d{\bQ}_2 \ldots d{\bQ}_{N-1} \, 
\exp \biggl[ - {H  \over 2  k_{\rm B} T}  \sum_{j = 2}^{N- 1} \bQ_j^2 
- i \left( \sum_{j = \mu}^{\nu - 1 } (1 - \delta_{1 j} ) \,  \bQ_j 
\right) \cdot \bfk  \nonumber \\ 
\nonumber \\ 
- {1 \over 2  k_{\rm B} T} \sum_{\alpha,\beta = 2 \atop \alpha \ne \beta}^N 
\, E \left( {\br}_{\alpha} - {\br}_{\beta} \right) \biggr] \,
\Biggl\lbrace \int \! \! d{\bQ}_1 \, 
\exp \biggl[ - {H \over 2 k_{\rm B} T } \, \bQ_1^2
- i  \, \delta_{1 \mu} \, \bQ_1 \cdot \bfk  \nonumber \\ 
\nonumber \\ 
- { 1 \over  k_{\rm B} T } \, \left[ E \left( {\br}_{1} - {\br}_{2} \right)
+ E \left( {\br}_{1} - {\br}_{3} \right) + \ldots 
+ E \left( {\br}_{1} - {\br}_{N} \right) \right]
\biggr] \Biggr\rbrace 
\label{int2} 
\end{eqnarray}
where, a typical term of the excluded volume potential contribution to 
the $\bQ_1$ integral, has the form,
\begin{eqnarray} 
E \left( {\br}_{1} - {\br}_{\beta} \right) &=& 
{v \, k_{\rm B} T \over [2 \pi {\tilde d}^2]^{3 \over 2}} \, 
\exp \Biggl\lbrace - {1 \over 2 {\tilde d}^2 } \, \biggl( \bQ_1^2 
+ 2 \, \bQ_1 \cdot \br_{\beta 2} + \bQ_2^2 + 2  \, \bQ_2 
\cdot \br_{\beta 3} + \ldots \nonumber \\ 
\nonumber \\ 
&+& \bQ_{\beta - 2}^2 + 2  \, \bQ_{\beta -2 } 
\cdot \br_{\beta, \, \beta -1 }  
+ \bQ_{\beta -1}^2 \biggr) \Biggr\rbrace \nonumber 
\label{evterm} 
\end{eqnarray}

We now convert the $\bQ_1$ integral into spherical coordinates. In
order to do so, we need to choose a reference vector to fix a
direction in space. In the $\bQ_1$ integration, all the other vectors,
$\bQ_2, \ldots, \bQ_{N-1}$ and $\bfk$ are fixed. Without loss of
generality, we choose the fixed vector as $\bQ_2$, denote its
direction as the $z$ direction, and choose, in the plane perpendicular
to $\bQ_2$, an arbitrary pair of orthogonal directions as the $x$ and
$y$ axes. Let, $\theta_1$, $\theta_{\beta 2}$, and $\theta_k$
represent the angles that the vectors $\bQ_1$, $\br_{\beta 2}$ and
$\bfk$ make with the $z$ direction, respectively. Similarly, let
$\phi_1$, $\phi_{\beta 2}$, and $\phi_k$ represent the angles that the
projections of these vectors on the $x y$ plane, make with the $x$
direction. Then,
\[
\bQ_1 \cdot \br_{\beta 2} = Q_1 \, r_{\beta 2} \, 
F_{\beta 2} \, (\theta_1, \phi_1) 
\]
where, $Q_1$ and $r_{\beta 2}$ represent the magnitudes of  $\bQ_1$ and 
$\br_{\beta 2}$, respectively, and,
\[ 
F_{\beta 2} \, (\theta_1, \phi_1) = \sin \theta_1 \sin \theta_{\beta 2} 
(\cos \phi_1 \cos \phi_{\beta 2} + \sin \phi_1 \sin \phi_{\beta 2}) 
+ \cos \theta_1 \cos \theta_{\beta 2} 
\]
Defining the function $F_k \, (\theta_1, \phi_1)$ similarly, we can 
rewrite the $\bQ_1$ integral in expression~\ref{int2}, in terms of 
spherical coordinates as, 
\begin{eqnarray} 
I_{Q_1} =  \int_0^\infty \! \! d Q_1 \int_0^{2 \pi} \! \! d \theta_1 
\int_0^{\pi} \! \! d \phi_1 \, Q_1^2 \, \sin \theta_1 \, 
\exp \left[ - {H  \over 2  k_{\rm B} T} \, Q_1^2  
- i  \, \delta_{1 \mu} \, Q_1 k \, F_k \, (\theta_1, \phi_1) 
\right] \times \nonumber \\ 
\nonumber \\ 
\exp \, \Biggl\lbrace - {1 \over k_{\rm B} T } \, \Biggl[
{v \, k_{\rm B} T \over \left(2 \pi {\tilde d}^2 \right)^{3 / 2}} \, 
\biggl\lbrace e^{ - {1 \over 2 {\tilde d}^2 } \, Q_1^2 } + 
e^{ - {1 \over 2 {\tilde d}^2 } \left( Q_1^2 
+ 2 \, Q_1 \, r_{3 2} \, F_{3 2} (\theta_1, \phi_1)  \, \right) } \,  
e^{ - {1 \over 2 {\tilde d}^2 } \, \bQ_2^2 } \nonumber \\ 
\nonumber \\ 
+ \ldots + e^{ - {1 \over 2 {\tilde d}^2 } \left( Q_1^2 + 2  \, Q_1 \, 
r_{N 2}  \, F_{N 2} (\theta_1, \phi_1)  \, \right) } \,  
e^{ - {1 \over 2 {\tilde d}^2 } \, \left( \bQ_2^2 + 
2  \, \bQ_2 \cdot \br_{N 3} + \ldots + \bQ_{N -1}^2 \right)} 
\biggr\rbrace \, \Biggr] \, \Biggr\rbrace 
\label{int3} 
\end{eqnarray}
For $Q_1=0$, the integrand is identically zero. For $Q_1 \ne 0$, 
in the limit ${\tilde d} \to 0$ or ${\tilde d} \to \infty$, 
the integrand tends to,
\[ 
Q_1^2 \, \sin \theta_1 \, 
\exp \left\lbrace - {H  \over 2  k_{\rm B} T} \, Q_1^2  
- i  \, \delta_{1 \mu} \, Q_1 k \, 
F_k \, (\theta_1, \phi_1) \right\rbrace
\]
The integrand is also a bounded function of $Q_1$ for all values 
${\tilde d}$. 

An argument similar to the one above can be carried out 
for each of the remaining integrations over $\bQ_2, \ldots, \bQ_{N-1}$. 
It follows that,
\begin{eqnarray} 
\lim_{{\tilde d} \to 0 \atop {\rm or}, \, {\tilde d} \to \infty} 
\int \! \! d{\bQ}_1 \ldots d{\bQ}_{N-1} \, 
\exp \left\lbrace - \left( {\phi \over k_{\rm B} T} \right) 
- i \left( \sum_{j = 
\mu}^{\nu - 1 } \bQ_j \right) \cdot \bfk \right\rbrace \nonumber \\
\nonumber \\
= \int \! \! d{\bQ}_1 \ldots d{\bQ}_{N-1} \, 
\exp \left\lbrace - \left( {H \over 2 k_{\rm B} T} \right) 
\sum_{j = 1}^{N- 1} \bQ_j^2 - i \left( \sum_{j = 
\mu}^{\nu - 1 } \bQ_j \right) \cdot \bfk \right\rbrace 
\label{int4} 
\end{eqnarray}
With regard to the normalization factor $\cal N_{\rm eq}$, since,
\begin{equation} 
{\cal N}_{\rm eq} = \left[ \int \! \! d{\bQ}_1 \ldots d{\bQ}_{N-1} \, 
\exp \left( - { \phi \over k_{\rm B} T} \right) \right]^{-1}
\label{int5} 
\end{equation} 
we can show, by adopting a procedure similar to that above that, 
\begin{equation} 
\lim_{{\tilde d} \to 0 \atop {\rm or}, \, {\tilde d} \to \infty} 
{\cal N}_{\rm eq} = {\cal N}_{\rm eq}^R 
\label{normlim} 
\end{equation}

As a result, we have established that, 
\begin{eqnarray}
\lim_{{\tilde d} \to 0 \atop {\rm or}, \, {\tilde d} \to \infty}  
P_{\rm eq} (\br_{\nu \mu}) &=& {1 \over (2 \pi)^3} \int \! \! d \bfk 
\, e^{i  \, \br_{\nu \mu} \cdot \bfk} \, \, 
\Biggl\lbrace \int \! \! d{\bQ}_1 \ldots d{\bQ}_{N-1} \, 
\psi_{\rm eq}^R  \, 
e^{ - \, i \left[ \sum_{j = \mu}^{\nu - 1 } 
\bQ_j \right] \cdot \bfk } \, \Biggr\rbrace \nonumber \\ 
\nonumber \\ 
&=& P_{\rm eq}^R (\br_{\nu \mu}) 
\label{int6} 
\end{eqnarray}

\section{Codeformational Memory-Integral Expansion}
\label{codef}

Upon expanding the tensors $\bsjk$ in the manner displayed in
eq~\ref{linvisexp}, substituting the expansion into the second moment
equation, eq~\ref{gaussecmom}, and separating the resultant equation
into equations for each order in the velocity gradient, the following
two equations are obtained, 
\vskip10pt
\noindent 
{\bf Equilibrium:}
\begin{equation}
{d  \over dt} \, f_{jk}  =
{ 2 k_{\rm B} T \over \zeta} \, A_{jk} 
- \left( { H \over \zeta} \right) \, \sum_{m=1}^{N-1}  \left[ 
f_{jm} \, (A_{mk} - {z^*} \, \Delta^{(0)}_{km} \, ) + 
(A_{jm} - {z^*} \, \Delta^{(0)}_{jm} \, ) \, f_{mk} 
\right] 
\label{equisecmom}
\end{equation}
where, 
\begin{equation}
\Delta^{(0)}_{jm} = \sum_{\mu=1}^{N}  \left[ \, 
{( B_{j+1,\, m} -  B_{\mu m}) \over \left( {d^*}^2 + {\hat f}_{j+1, \, \mu} \right)^{5/2}}
- {( B_{j m} -  B_{\mu m}) \over \left( {d^*}^2 + {\hat f}_{j \mu} \right)^{5/2}}
\, \right]
\label{Del0}
\end{equation}
with the quantities ${\hat f}_{\nu \mu}$ given by, 
\begin{equation}
{\hat f}_{\nu \mu} = \left( {H \over k_B T } \right)
\, \sum_{j,k \,= \,\min(\mu,\nu)}^{ \max(\mu,\nu)-1} \, f_{jk}
\end{equation}
\vskip10pt
\noindent 
{\bf First Order:}
\begin{equation}
{d \over dt} \, \beps_{jk} = 
( \bk + \bk^T ) \, f_{jk} -  \left( { H \over \zeta} \right) \, 
\sum_{m, n =1}^{N-1} {\overline A}_{jk, \, mn} \, \beps_{m n} 
\label{fordsecmom}
\end{equation}
where,  
\begin{eqnarray}
{\overline A}_{jk, \, mn} = ( A_{jm} \, \delta_{kn} + \delta_{jm}  \, A_{kn})
- {z^*} \, ( \Delta^{(0)}_{jm} \, \delta_{kn} + \delta_{jm}  
\, \Delta^{(0)}_{kn} ) \nonumber \\
\nonumber \\
+ {z^*} \, \left( { H \over k_B T } \right) \, \sum_{p=1}^{N-1}  
\left[ f_{jp} \, \Delta^{(1)}_{kp, \, mn} 
+  \Delta^{(1)}_{jp, \, mn} \, f_{pk} \right]
\label{abar}
\end{eqnarray}
with the quantities $ \Delta^{(1)}_{jp, \, mn} $ given by,
\begin{equation}
\Delta^{(1)}_{jp, \, mn} = \sum_{\mu=1}^{N}  \left[ \, 
{( B_{j+1,\, p} -  B_{\mu p}) \, \theta (\mu,m,n,j+1)
\over \left( {d^*}^2 + {\hat f}_{j+1, \, \mu} \right)^{7/2}}
- {( B_{j p} -  B_{\mu p}) \, \theta (\mu, m,n,j) 
\over \left( {d^*}^2 + {\hat f}_{j \mu} \right)^{7/2}} \, \right]
\label{Del1}
\end{equation}
The function $\theta(\mu,m,n,\nu)$ has been introduced previously in the 
treatment of hydrodynamic interaction~\cite{ottga}. It is unity if $m$ and $n$ lie
between $\mu$ and $\nu$, and zero otherwise,
\begin{equation}
\theta(\mu,m,n,\nu)=\cases{1& if $\mu \leq m,n < \nu$ \quad or\quad
$\nu \leq m,n < \mu $\cr
\noalign{\vskip3pt}
0& otherwise\cr} 
\label{theta}
\end{equation}
Introducing new indices for the pairs $(j,k)$ and $(m,n)$, the 
quantity ${\overline A}_{jk, \, mn}$ may be considered an
$(N-1)^2 \times (N-1)^2$ matrix. The inverse can then be defined 
in the following manner, 
\begin{equation}
\sum_{r, s=1}^{N-1} {\overline A}^{\, -1}_{jk, \, rs} 
\, {\overline A}_{rs, \, mn} 
= \one_{ jk, \, mn}
\end{equation}
where,  $\one$ is the $(N-1)^2 \times (N-1)^2$ unit matrix 
$\one_{ jk, \, mn} = \delta_{jm} \, \delta_{kn}$. 

In the equilibrium second moment equation, eq~\ref{equisecmom}, 
the term $(d f_{jk} / dt) $ on the left hand side, is identically zero. It 
is retained here, however, to indicate that the equation is solved for 
$f_{jk} $ by numerical integration of the ODE's until steady state is reached. 

Upon integrating eq~\ref{fordsecmom} with respect to time, 
and substituting the solution into eq~\ref{gausskram}, 
we finally obtain the expression, eq~\ref{memexp}, for the 
first order codeformational memory-integral expansion, 
where, the memory function $G(t)$ is given by, 
\begin{equation}
G(t)=\sum_{j,k=1}^{N-1} \,  \sum_{m,n=1}^{N-1} f_{jm} \, {\cal H}_{jk} \,
{\cal E}  \biggl\lbrack \,  - \, {t \over \lambda_H} \, {\widetilde A}  
\, \biggr\rbrack_{\, mk, \, nn} 
\label{memfun}
\end{equation}
In eq~\ref{memfun}, $\lambda_H = (\zeta / 4H)$ is the familiar time 
constant, ${\cal H}_{jk}$ is defined by,
\begin{equation}
{\cal H}_{jk} = n_{\rm p} H   \left[ \delta_{jk} - {1 \over 2} \, {z^*} 
\sum_{\mu ,\nu=1 \atop \mu \ne \nu }^{N} { {d^*}^2 
\over \left( {d^*}^2 + {\hat f}_{\nu \mu} \right)^{7/2}} 
\, \theta (\nu,j,k,\mu) \right]
\label{calh}
\end{equation}
and the quantity ${\widetilde A}_{jk, \, mn}$ is given by, 
\begin{equation}
{\widetilde A}_{jk, \, mn} = \sum_{r, s=1}^{N-1} {1 \over 4} \, 
f^{-1}_{jr} \, {\overline A}_{rk, \, sn} \, f_{sm}
\end{equation}
The exponential operator $ {\cal E} \, [\, M \,]$ maps one matrix into another 
according to: 
\[ {\cal E} \, [\, M  \, ]_{\, jk, \, mn}=\one_{jk, \, mn} + M_{jk, \, mn} + {1 \over 2!} 
\sum_{r, s=1}^{N-1} \, M_{jk, \, rs } \, 
M_{rs, \, m n} + \ldots \]
and has the useful properties,
\[ {d \over dt} \, {\cal E} \, [\, M t \, ]_{\, jk, \, mn }=  \sum_{r, s=1}^{N-1} 
M_{jk, \, rs} \, {\cal E} \, [\, M t \, ]_{\, rs, \, mn} 
=\sum_{r, s=1}^{N-1}  {\cal E} \, [ \, Mt \, ]_{\, jk, \, rs} \,  M_{rs, \, mn} \] 
\[ \sum_{r, s=1}^{N-1}  {\cal E} \, [\, a \, M  \, ]_{\, jk, \, rs } \, \, 
{\cal E} \, [\, b \, \one  \, ]_{\, rs, \, mn} 
={\cal E} \, [ \, a \, M + b \, \one \, ]_{\, jk, \, mn } \] 
for arbitrary constants $a$ and $b$. 

As mentioned in section~5, once the memory function $G(s)$ is
obtained, one can obtain the material functions in small amplitude
oscillatory shear flow, and the zero shear rate viscometric
functions. Following the procedure outlined in section~5, we can show
that, 
\begin{eqnarray}
\eta^\prime (\omega) &=& \lambda_H \sum_{j,k=1}^{N-1} \,  
\sum_{m,n =1}^{N-1}  \,    \sum_{r,s =1}^{N-1}   
f_{jk} \, {\cal H}_{jm} \,
\Phi^{-1}_{k m, \, r s} \, {\widetilde A}_{rs, \, nn} \nonumber \\
\nonumber \\
\eta^{\prime \prime} (\omega)  &=& \lambda_H^2  \, 
\omega \sum_{j,k=1}^{N-1} \,  
\sum_{m,n =1}^{N-1}  
f_{jk} \, {\cal H}_{jm} \,
\Phi^{-1}_{k m, \, nn} 
\label{etaprime}
\end{eqnarray}
where,
\begin{equation} 
\Phi_{jk, \, mn} =  \sum_{r,s =1}^{N-1}  
\left[ {\widetilde A}_{jk, \, rs} \, {\widetilde A}_{rs, \, mn} 
+ \lambda_H^2  \, \omega^2 \one_{jk, \, mn} \right]
\end{equation} 
Using the relations between the zero shear rate viscometric functions 
and $\eta^{\prime}$ and $\eta^{\prime\prime}$ (eqs~\ref{usf8}), one
can show that,
\begin{eqnarray}
\eta_{p,0} &=& 4 \, \lambda_H \sum_{j,k=1}^{N-1} \,  
\sum_{m,n =1}^{N-1}  
{\cal H}_{jk} \,
{\overline A}^{\, -1}_{jk, \, m n} \, f_{mn} 
\label{stetap0} \\
\nonumber \\
\Psi_{1,0} &=& 32 \, \lambda_H^2  \sum_{j,k=1}^{N-1} \,  
\sum_{m,n =1}^{N-1} \,  \sum_{r,s =1}^{N-1} 
{\cal H}_{jk} \, {\overline A}^{\, -1}_{jk, \, m n} \, 
{\overline A}^{\, -1}_{mn, \, rs} \, f_{rs}
\label{stpsi10}
\end{eqnarray}

\section{Viscometric functions correct to first order in $z^*$}
\label{visco}

The first step in calculating the first order excluded volume
corrections to the Rouse viscometric functions, as mentioned earlier,
is to evaluate the time integrals in eqs~\ref{qqtheta}
and~\ref{pertau}.  These time integrals can be carried 
out by using the forms of the tensors ${\bgam}_{[1]}$ and
$\bE$ in steady shear flow, tabulated in reference~12. 
One can show that the expression for
the second moment $\bs^R_{jk}$, which is required to explicitly
evaluate all the averages carried out with the Rouse distribution function
$\psi_R$, is given by,
\begin{equation}
\bs^R_{jk} =  {k_{\rm B} T \over H} \left\lbrace 
\delta_{jk} \, \bu + 2 \lambda_H C_{jk} \left( \bk 
+\bk^T \right) + 8 \lambda_H^2 C^2_{jk} \left( \bk \cdot \bk^T \right) 
\right\rbrace
\label{bsrjk}
\end{equation}
while the stress tensor in steady shear flow has the form,
\begin{eqnarray} 
\btau^p &=& - n_{\rm p} k_BT  \sum_{j =1}^{N-1} 
\left[ 2 \lambda_H \, C_{jj} \left( \bk +\bk^T \right) 
+ 8 \lambda_H^2 \, C^2_{jj} 
\left( \bk \cdot \bk^T \right) \right] \nonumber \\ 
\nonumber \\ 
&-& n_{\rm p} H  \sum_{j,k =1}^{N-1} 
\biggl\lbrace  2 \lambda_H \, C_{kj} \, \bY^R_{jk}
+ 4 \lambda_H^2 \, C^2_{kj}  \left[ \bk \cdot \bY^R_{jk} 
+ \bY^R_{jk} \cdot \bk^T \right] \nonumber \\
\nonumber \\ 
&+& 16 \lambda_H^3 \, C^3_{kj}  \left[ \bk \cdot \bY^R_{jk} \cdot
\bk^T \right] \biggr\rbrace + \bZ^R
\label{stress1}
\end{eqnarray}
A similar expression, without the $\bZ^R$ term, has been derived by 
\"Ottinger~\cite{ottrg} in his renormalisation group treatment of 
excluded volume effects---within the framework of polymer kinetic 
theory---using a $\delta$-function excluded volume potential.  

\begin{table}[bt!]
\caption{ Functions, appearing in eq~\ref{stress2}, representing the  
{\em indirect}  excluded volume contributions to the stress tensor 
in steady shear flow. The quantity $\Omega$ is defined by, $\Omega =
4 / ( {d^*}^2 + S_{\mu \nu}^{(0)} )^2$.}
\begin{tabular*}{\textwidth}{@{}l@{\extracolsep{\fill}}ll}
& & \\
\hline
& & \\
{  $ \alpha^{(0)}_{\mu \nu} = \lambda_H  \, S_{\mu \nu}^{(0)}
\, \Omega 
\left[  \left( {d^*}^2 + S_{\mu \nu}^{(0)} \right)^2 
+  \lambda_H^2 {\dot \gamma}^2 \, \left\lbrace 2 \, S_{\mu \nu}^{(2)} 
\left( {d^*}^2 + S_{\mu \nu}^{(0)} \right)
- S_{\mu \nu}^{(1)} \, S_{\mu \nu}^{(1)} \right\rbrace \right] $ } \\
& & \\
& & \\
{     $ \alpha^{(1)}_{\mu \nu} = 
\lambda_H^2  \, S_{\mu \nu}^{(1)} \, \Omega 
\left[  \left( {d^*}^2 + S_{\mu \nu}^{(0)} \right)  
\left( 2 {d^*}^2 + S_{\mu \nu}^{(0)} \right) 
+ 2 \lambda_H^2 {\dot \gamma}^2 \left\lbrace 2 \, S_{\mu \nu}^{(2)} 
\left( {d^*}^2 + S_{\mu \nu}^{(0)} \right) 
-S_{\mu \nu}^{(1)} \, S_{\mu \nu}^{(1)} \right\rbrace \right] $ } \\
& & \\
& & \\
{   $ \alpha^{(2)}_{\mu \nu} = \lambda_H^3 
\, \Omega
\left[ \, 3 \, {d^*}^2 + 2 \, S_{\mu \nu}^{(0)}
+ 6 \lambda_H^2 {\dot \gamma}^2 \, S_{\mu \nu}^{(2)} 
+ \, 2  \, S_{\mu \nu}^{(2)} \left( {d^*}^2 + S_{\mu \nu}^{(0)} \right) 
- S_{\mu \nu}^{(1)} \, S_{\mu \nu}^{(1)} 
\right] $ } \\
& & \\
& & \\
{   $ \alpha^{(3)}_{\mu \nu} = - \lambda_H^3 
\, \Omega \, 
S_{\mu \nu}^{(1)} \, S_{\mu \nu}^{(1)}  \, {d^*}^2 $ } \\
& & \\
& & \\
{   $ \alpha^{(4)}_{\mu \nu} = \lambda_H^4  \, S_{\mu \nu}^{(1)}
\, \Omega
\left[  \, 2 \, S_{\mu \nu}^{(1)} \, S_{\mu \nu}^{(1)} - 3 \, 
S_{\mu \nu}^{(2)} \left( {d^*}^2 + S_{\mu \nu}^{(0)} \right) \right] $ } \\
& & \\
& & \\
{   $ \alpha^{(5)}_{\mu \nu} = \alpha^{(4)}_{\mu \nu} $}  \\
& & \\
& & \\
{   $ \alpha^{(6)}_{\mu \nu} = 2 \, \lambda_H^5  
\, \Omega
\left[  \, 3  \, S_{\mu \nu}^{(2)} \left\lbrace S_{\mu \nu}^{(1)} \, S_{\mu \nu}^{(1)} 
- S_{\mu \nu}^{(2)} \left( {d^*}^2 + S_{\mu \nu}^{(0)} \right) \right\rbrace 
- 2 \, S_{\mu \nu}^{(3)} \, S_{\mu \nu}^{(1)}  \left( {d^*}^2 + 
S_{\mu \nu}^{(0)} \right) \right] $ } \\
& & \\
& & \\
\hline
\end{tabular*}
\vskip10pt
\end{table}

\begin{table}[!bt]
\caption{ Functions, appearing in eq~\ref{stress2},
representing the {\em direct}  excluded volume 
contributions to the stress tensor 
in steady shear flow. The quantity $\Omega$ is defined by,
$\Omega = 4 / ( {d^*}^2 + S_{\mu \nu}^{(0)} )^2$. }
\begin{tabular*}{\textwidth}{@{}l@{\extracolsep{\fill}}ll}
& & \\
\hline
& & \\
{   $ \beta^{(0)}_{\mu \nu} = \lambda_H  \, S_{\mu \nu}^{(0)}
\, \Omega
\left[  \left( {d^*}^2 + S_{\mu \nu}^{(0)} \right)^2 
+  \lambda_H^2 {\dot \gamma}^2 \, \left\lbrace 2 \, S_{\mu \nu}^{(2)} 
\left( {d^*}^2 + S_{\mu \nu}^{(0)} \right)
- S_{\mu \nu}^{(1)} \, S_{\mu \nu}^{(1)} \right\rbrace \right] $ } \\
& & \\
& & \\
{   $ \beta^{(1)}_{\mu \nu} = \lambda_H^2  \, S_{\mu \nu}^{(1)}
\, \Omega
\left[  \left( {d^*}^2 + S_{\mu \nu}^{(0)} \right)  
 {d^*}^2 +  \lambda_H^2 {\dot \gamma}^2 \left\lbrace 2 \, S_{\mu \nu}^{(2)} 
\left( {d^*}^2 + S_{\mu \nu}^{(0)} \right) 
-S_{\mu \nu}^{(1)} \, S_{\mu \nu}^{(1)} \right\rbrace \right] $ } \\
& & \\
& & \\
{   $ \beta^{(2)}_{\mu \nu} = \lambda_H^3 
\, \Omega
\left[ \, {d^*}^2 + 2 \lambda_H^2 {\dot \gamma}^2 \, S_{\mu \nu}^{(2)} 
+ \, 2  \, S_{\mu \nu}^{(2)} \left( {d^*}^2 + S_{\mu \nu}^{(0)} \right) 
- S_{\mu \nu}^{(1)} \, S_{\mu \nu}^{(1)} 
\right] $ } \\
& & \\
& & \\
{   $ \beta^{(3)}_{\mu \nu} = - \lambda_H^3 
\, \Omega \, 
S_{\mu \nu}^{(1)} \, S_{\mu \nu}^{(1)}  \, {d^*}^2 $ } \\
& & \\
& & \\
{   $ \beta^{(4)}_{\mu \nu} = \lambda_H^4  \, S_{\mu \nu}^{(1)}
\, \Omega
\left[  \, S_{\mu \nu}^{(1)} \, S_{\mu \nu}^{(1)} - 2 \, 
S_{\mu \nu}^{(2)} \left( {d^*}^2 + S_{\mu \nu}^{(0)} \right) \right] $ } \\
& & \\
& & \\
{   $ \beta^{(5)}_{\mu \nu} = \beta^{(4)}_{\mu \nu} $} \\
& & \\
& & \\
{   $ \beta^{(6)}_{\mu \nu} = 2 \,  \lambda_H^5  
\, \Omega
\left[  \, S_{\mu \nu}^{(1)} \, S_{\mu \nu}^{(1)} 
- 2 \, S_{\mu \nu}^{(2)} \left( {d^*}^2 + S_{\mu \nu}^{(0)} \right)
\right] $ } \\ 
& & \\
& & \\
\hline
\end{tabular*}
\vskip10pt
\end{table}

The next step is to explicitly evaluate the tensors $\bY_{jk}^R$ and 
$\bZ^R$, in terms of the velocity gradient $\bk$ and the Kramers matrix
$C_{kj}$, by using eq~\ref{bsrjk} for $\bs^R_{jk}$. The resultant 
expressions are then substituted into eq~\ref{stress1}, and 
after a lengthy calculation, the following explicit expression for the 
{\em excluded volume contribution} to the stress tensor, correct 
to first order in $z^*$, is obtained, 
\begin{eqnarray} 
\btau^p_{(E)} &=& - {1 \over 8 \lambda_H } \, n_{\rm p} k_BT \, z^*  
\sum_{\mu, \nu =1 \atop \mu \ne \nu}^{N}
{1 \over \left( {d^*}^2 + S_{\mu \nu}^{(0)} \right)^{5/2} 
\, e_{\mu \nu}({\dot \gamma})^{3/2} } \, 
\Biggl\lbrace \left( \alpha_{\mu \nu}^{(0)} - \beta_{\mu \nu}^{(0)}  
\right) \bu \nonumber \\
\nonumber \\
&+&  \left( \alpha_{\mu \nu}^{(1)} - \beta_{\mu \nu}^{(1)} \right) 
\left( \bk + \bk^T \right) 
+  \left( \alpha_{\mu \nu}^{(2)} - \beta_{\mu \nu}^{(2)} \right) 
\left( \bk \cdot \bk^T \right)  
+  \left( \alpha_{\mu \nu}^{(3)} - \beta_{\mu \nu}^{(3)} \right) 
\left( \bk^T \cdot \bk \right)  \nonumber \\
\nonumber \\
&+&  \left( \alpha_{\mu \nu}^{(4)} - \beta_{\mu \nu}^{(4)} \right) 
\left( \bk \cdot \bk^T \cdot \bk \right) 
+  \left( \alpha_{\mu \nu}^{(5)} - \beta_{\mu \nu}^{(5)} \right) 
\left( \bk^T \cdot \bk \cdot \bk^T \right)  \nonumber \\
\nonumber \\
&+&  \left( \alpha_{\mu \nu}^{(6)} - \beta_{\mu \nu}^{(6)} \right) 
\left( \bk \cdot \bk^T \cdot \bk \cdot \bk^T \right) \Biggr\rbrace
\label{stress2}
\end{eqnarray}
where, the functions $\alpha_{\mu \nu}^{(j)}$ and $\beta_{\mu
\nu}^{(j)}; \, (j=0,1, \ldots, 6)$, which represent the {\em indirect}
and {\em direct} contributions respectively, are given in
Tables~1 and~2, and the function $e_{\mu \nu}({\dot
\gamma})$ is defined by,
\begin{equation}
e_{\mu \nu}({\dot \gamma}) = 1 + 
{ \lambda_H^2 {\dot \gamma}^2 
\over \left( {d^*}^2 + S_{\mu \nu}^{(0)} \right)^2 } 
\left[ 2 \left( {d^*}^2 + S_{\mu \nu}^{(0)} \right) S_{\mu \nu}^{(2)}
- S_{\mu \nu}^{(1)} \, S_{\mu \nu}^{(1)} \right]
\label{emunu}
\end{equation}
Equation~\ref{stress2} for the stress tensor can then be used to find
the {\em excluded volume contributions} to  
the viscometric functions, correct to first order in $z^*$, 
by using the definitions in eqs~\ref{sfvis},
\begin{eqnarray} 
\eta_{p}^{(E)} &=&  {1 \over 2}  \, \lambda_H^2 \, z^*  
\sum_{\mu, \nu =1 \atop \mu \ne \nu}^{N}
{1 \over \left( {d^*}^2 + S_{\mu \nu}^{(0)} \right)^{7/2} 
e_{\mu \nu}({\dot \gamma})^{3/2} } \, 
\biggl[ \, S_{\mu \nu}^{(0)} \, S_{\mu \nu}^{(1)}  
+  \lambda_H^2 {\dot \gamma}^2  \, S_{\mu \nu}^{(1)}  S_{\mu \nu}^{(2)} \nonumber \\
&+& {d^*}^2 \, S_{\mu \nu}^{(1)} \, \biggr]
\label{pertetap} \\ 
\nonumber \\
\Psi_{1}^{(E)} &=&   \lambda_H^2 \, z^*  
\sum_{\mu, \nu =1 \atop \mu \ne \nu}^{N}
{1 \over \left( {d^*}^2 + S_{\mu \nu}^{(0)} \right)^{7/2} 
\, e_{\mu \nu}({\dot \gamma})^{3/2} } \nonumber \\
\nonumber \\
&\times& \left[ \, 2 \, S_{\mu \nu}^{(2)} \left( {d^*}^2 + S_{\mu \nu}^{(0)} \right)
- S_{\mu \nu}^{(1)} \, S_{\mu \nu}^{(1)} 
+  \lambda_H^2 {\dot \gamma}^2 \, \left( 3 \, S_{\mu \nu}^{(2)} \, S_{\mu \nu}^{(2)} 
- 2\, S_{\mu \nu}^{(3)} \, S_{\mu \nu}^{(1)} \right) \right]
\label{pertpsi1} \\
\nonumber \\
\Psi_{2}^{(E)} &=&  {1 \over 8 \lambda_H } \, n_{\rm p} k_BT \, z^*  
\sum_{\mu, \nu =1 \atop \mu \ne \nu}^{N}
{\alpha^{(3)}_{\mu \nu} - \beta^{(3)}_{\mu \nu} \over 
\left( {d^*}^2 + S_{\mu \nu}^{(0)} \right)^{5/2} 
\, e_{\mu \nu}({\dot \gamma})^{3/2} }  = 0
\label{pertpsi2}  
\end{eqnarray}  
These expressions have been derived earlier by
\"Ottinger~\cite{ottrg}, in an arbitrary number of space dimensions,
in the limit $d^* \to 0$. It is clear from eq~\ref{pertpsi2} that the
second normal stress difference coefficient is zero because the
indirect and direct excluded volume contributions cancel each other
out.  When $d^* \to 0$, however, both the quantities
$\alpha^{(3)}_{\mu \nu}$ and $ \beta^{(3)}_{\mu \nu} $ are identically
zero.


\begin{thebibliography}{99}


\bibitem{rouse} Rouse, P.E.; {\it J. Chem. Phys.} {\bf 1953}, 21, 1272.  

\bibitem{ottga} \"{O}ttinger, H. C. {\it J. Chem. Phys.} {\bf 1989}, 90, 463.

\bibitem{wedgega} Wedgewood, L. E. {\it J. Non-Newtonian Fluid Mech.} 
{\bf 1989}, 31, 127.

\bibitem{zylkaga} Zylka, W. {\it J. Chem. Phys.} {\bf 1991}, 94, 4628.

\bibitem{prakotthi} Prakash, J. R.; \"{O}ttinger, H. C. {\it J.
Non-Newtonian Fluid Mech.} {\bf 1997}, 71, 245.

\bibitem{prakbk}
Prakash, J. R. `The Kinetic Theory of Dilute Solutions of Flexible 
Polymers: Hydrodynamic Interaction'; In {\it Advances in the Flow 
and Rheology of Non-Newtonian Fluids}; Siginer, D. A.; Kee, D. De; 
Chhabra, R. P.; Eds; Rheology Series; Elsevier Science: Amsterdam, 1999.

\bibitem{prakottev} Prakash, J. R.; \"{O}ttinger, H. C. {\it Macromolecules} 
{\bf 1999}, 32, 2028.

\bibitem{doi} Doi, M.; Edwards, S. F. {\it The Theory of Polymer
Dynamics}; Oxford University Press: Oxford, 1986.

\bibitem{declos} des Cloizeaux, J.; Jannink, G. {\it Polymers in
Solution, Their Modelling and Structure}; Oxford University Press: 
Oxford, 1990.

\bibitem{schaf} Sch\"{a}fer, L. {\it Excluded Volume Effects in Polymer 
Solutions};  Springer: Berlin, 1999.

\bibitem{bird2} Bird, R. B.; Curtiss, C. F.; Armstrong, R. C.; 
Hassager, O.  {\it Dynamics of Polymeric Liquids. Kinetic Theory,}
2nd edn.; Wiley-Interscience: New York, 1987; Vol.~2.

\bibitem{bird1} Bird, R. B.; Armstrong, R. C.; Hassager, O.  {\it Dynamics
of Polymeric Liquids. Fluid Mechanics,} 2nd edn.; Wiley-Interscience: 
New York, 1987; Vol.~1.

\bibitem{ottbk} \"{O}ttinger, H. C. {\it Stochastic Processes in
Polymeric Fluids};  Springer: Berlin, 1996.
 
\bibitem{schiebiv} Schieber, J. D.; {\it J. Rheol.} {\bf 1993}, 37, 1003.

\bibitem{wedgeiv} Wedgewood, L. E. {\it Rheol. Acta} {\bf 1993}, 32, 405.

\bibitem{ottrabrg} \"{O}ttinger, H. C.; Rabin, Y. {\it J. 
Non-Newtonian Fluid Mech.} {\bf 1989}, 33, 53.

\bibitem{ottrg} \"{O}ttinger, H. C. {\it Phys. Rev.} {\bf 1989}, A40,
2664.

\bibitem{zylkarg} Zylka, W.; \"{O}ttinger H. C. {\it Macromolecules}
{\bf 1991}, 24, 484.

\bibitem{numrec} Press, W. H.;  Teukolsky, S. A.; Vetterling, W. T.; 
Flannery, B.  P. {\it Numerical Recipes in FORTRAN}, 2nd edn.; Cambridge
University Press: Cambridge, 1992.

\bibitem{rem1} The scaling of linear viscoelastic properties with 
molecular weight, and the frequency dependence of oscillatory shear 
flow material functions, for instance, seems to be nearly entirely 
determined by hydrodynamic interaction effects. 

\end{thebibliography}
\end{document}